\documentclass{aa}  
\usepackage[dvipsnames]{xcolor}
\usepackage{hyperref}
\usepackage{graphicx}
\usepackage{txfonts}
\usepackage{lipsum}
\usepackage{subcaption}         % necessary for continued figures, example in section 3
\usepackage{lscape}             % to rotate a single page table, example in appendix.
\usepackage{placeins}           % useful with \FloatBarrier, to keep 
\let\linenumbers\nolinenumbers
                                
\begin{document}

%%%%%%%%%%%%%%%%%%%%%%%%%%%%%%%%%%%%%%%%
% if you use custom commands in your title,
% ensure to check your title when submitting!
%%%%%%%%%%%%%%%%%%%%%%%%%%%%%%%%%%%%%%%%
   \title{Optical-morphology-based assessment of astrometric quality in Gaia-CRF3 quasars}

%   \subtitle{Subtitle}

%%%%%%%%%%%%%%%%%%%%%%%%%%%%%%%%%%%%%%%%
% Please separate each author with the \and command
%
% Please do not include ORCIDs next to author names.
% Only ORCIDs authenticated by individual authors in EDPS
% editorial system will be taken into account.
% ORCIDs included here will be removed.
%%%%%%%%%%%%%%%%%%%%%%%%%%%%%%%%%%%%%%%%

   \author{Qiqi Wu\inst{1}
        \and Shilong Liao\inst{1,2}\fnmsep\thanks{Corresponding author, \email{shilongliao@shao.ac.cn}}
        \and Zhaoxiang Qi\inst{1,2}\fnmsep\thanks{Corresponding author, \email{zxqi@shao.ac.cn}}
        \and Qi Xu\inst{1,2}
        \and Ye Ding\inst{1,2}
        \and Keyu Zhu\inst{1,2}
        }

   \institute{Shanghai Astronomical Observatory, Chinese Academy of Sciences, Shanghai 200030, China.
         \and{University of Chinese Academy of Sciences, Beijing 100049, China.}}

   \date{Received September 30, 20XX}

% \abstract{}{}{}{}{}
% 5 {} token are mandatory
 
  \abstract
   {Several studies have shown that host-galaxy structure or extended optical morphology in AGNs can induce spurious parallaxes and proper motions in Gaia DR3. However, it remains unclear whether source morphology also introduces systematic errors into the celestial reference frame constructed from Gaia data.}
   {We aim to provide a Gaia-independent external morphological indicator for Gaia-CRF3 sources and to use it to quantify the astrometric systematics associated with source morphology.} 
   {Using morphological parameters derived from DESI, SDSS, and SkyMapper, together with the PS1-PSC point-source score as a common reference scale, we used XGBoost to infer external morphological scores for Gaia-CRF3 sources. We then developed a multi-survey fusion scheme to combine the four survey-based point-source scores into a single composite score that measures the degree to which each source departs from the morphology of an ideal point source.}
   {We obtained morphological scores for 1,607,490 Gaia-CRF3 sources, corresponding to a completeness of 99.59\% with respect to the full Gaia-CRF3 catalogue. The score ranges from 0 to 1 and remains reliable for sources with $G<20.85$ mag. Based on this indicator, we find that AGNs with strongly non-point-like morphology induce a parallax zero-point shift of about $-43.7\,\mu$as, which cannot be effectively removed by the current parallax zero-point correction model. We also find that reference-source subsamples selected in different score ranges exhibit significantly different all-sky proper motion fields. For the high-purity point-source subsample with \texttt{point\_score} > 0.95, the total frame spin amplitude is reduced by 15.8\% relative to that of the full Gaia-CRF3 sample.}
   {The morphological score presented in this work provides an effective tracer of the optical morphology of Gaia-CRF3 sources. Morphologically anomalous sources in Gaia-CRF3 introduce distinct astrometric systematics and should therefore be removed or explicitly corrected in future constructions of higher-precision celestial reference frames.}


   \maketitle

%%%%%%%%%%%%%%%%%%%%%%%%%%%%%%%%%%%%%%%%%%%%%%%%%%%%%%%%%%%%%%
\section{Introduction}
As the backbone of fundamental astronomy, the Celestial Reference Frame (CRF) provides the inertial coordinate system required for high-precision astrometry, deep-space navigation, and geodetic monitoring. In 1998, the International Astronomical Union (IAU) officially adopted the International Celestial Reference Frame (ICRF) to replace the FK5 as the fundamental coordinate frame for astronomy \citep{1998AJ....116..516M}. This marked the first time that extragalactic sources, rather than Galactic stars, were used as fiducial markers, establishing a truly quasi-inertial reference frame. Over the past decades, the ICRF has evolved through successive realizations, progressing from ICRF1 to the current ICRF3 \citep{2015AJ....150...58F,2020A&A...644A.159C}. These realizations relied primarily on Very Long Baseline Interferometry (VLBI) observations of hundreds to thousands of radio quasars. While VLBI achieves exceptional precision in the radio domain, the ICRF has historically faced challenges regarding sparse sample density and uneven coverage, particularly in the southern hemisphere. The advent of the Gaia mission \citep{2016A&A...595A...1G} marked a shift from radio to optical astrometry and from sparse samples to massive datasets. Gaia-CRF3 \citep{2022A&A...667A.148G}, constructed using approximately 1.6 million AGNs (or candidates), represents the most precise all-sky optical reference frame currently available, increasing the density of reference sources by orders of magnitude while achieving alignment with ICRF3 at the micro-arcsecond ($\mu\text{as}$) level. This unprecedented sample density offers a novel opportunity to define an inertial system with a much more rigid and dense grid than ever before.

Extragalactic AGNs are expected to have negligible intrinsic proper motions and parallaxes, and are therefore natural fiducial objects for the construction of an optical celestial reference frame \citep{2018A&A...616A..14G, 2022A&A...667A.148G}. In practice, however, many AGNs in Gaia exhibit significant apparent proper motions, parallaxes, or degraded astrometric solutions \citep{2022ApJ...933...28M}. The physical origin of these signals is not always clear. In addition to residual calibration effects and limitations of the astrometric modelling \citep{2021A&A...649A...4L}, one plausible source-dependent contribution is that the optical source observed by Gaia does not behave as an ideal point source. Extended host-galaxy, asymmetric optical structure, foreground contamination, dual AGNs, and gravitationally lensed quasars can all destabilise the measured photocentre \citep{2017MNRAS.471.3775P, 2020ApJ...888...73H, 2023A&A...674A..11D}. 

The importance of this issue is reinforced by the systematic errors already identified in Gaia-CRF3. The Gaia team reported a global parallax zero-point offset of $-17~\mu$as \citep{2021A&A...649A...2L}, while subsequent studies showed that the parallax zero point depends on magnitude, colour, and ecliptic latitude \citep{2021A&A...649A...4L}. Statistical analyses of quasars and quasar candidates further revealed angular-scale-dependent systematics in both parallax and proper motion. On angular scales larger than 10 deg, the rms amplitudes of systematic errors in parallax and in a single proper motion component are approximately $8.1~\mu$as and $7.7~\mu$as yr$^{-1}$, respectively, increasing to $14.3~\mu$as and $17.2~\mu$as yr$^{-1}$ on angular scales larger than 0.5 deg \citep{2021A&A...649A...2L}. The inferred parallax zero point also depends on the adopted sample and method, ranging from about $-21~\mu$as for the full five-parameter sample to $-27~\mu$as for six-parameter sources, and reaching even larger values near the Galactic plane \citep{2021PASP..133i4501L,2024A&A...691A..81D}. These results demonstrate that the astrometric systematics of Gaia quasars cannot be characterised by a single global correction alone.

The key question is therefore whether part of these systematics is related to the intrinsic or environmental properties of the sources themselves. Previous studies have mainly addressed the purity of Gaia-CRF3, in particular contamination by stars and galaxies \citep{2022A&A...667A.148G,2023A&A...674A..41G,2023RAA....23b5006W}. However, even sources securely identified as AGNs are not guaranteed to be optimal reference-frame objects at milliarcsecond precision. Our previous work showed that spectroscopically confirmed AGNs can display anomalous proper motions and parallaxes, and that some of the most extreme cases include astrophysically interesting systems such as dual AGNs and lensed quasars \citep{2022FrASS...922768W,2024A&A...692A.154W}. This suggests that excluding only the most obvious outliers is insufficient. What is needed is a quantitative assessment of how the astrometric quality of Gaia-CRF3 sources varies with source properties, and in particular with optical morphology.

The importance of source morphology for reference frame construction is already well established in the radio domain. In the construction of ICRF3, the defining sources were selected not only for positional stability and sky distribution, but also for compact or quasi-point-like VLBI structure, while the source structure index was used as an indicator of astrometric suitability \citep{1997ApJS..111...95F,2015AJ....150...58F,2020A&A...644A.159C}. This suggests that morphology should also be examined carefully in the optical domain, where extended host-galaxy light or asymmetric structure can likewise displace the measured photocentre. Gaia provides several diagnostics that are sensitive to this behaviour, including \texttt{ipd\_frac\_multi\_peak}, which indicates the presence of multiple peaks in the image-parameter determination process \citep{2022NatAs...6.1185M}, \texttt{astrometric\_excess\_noise}, which reflects unmodelled residuals in the astrometric solution \citep{2020ApJ...888...73H}, and the renormalised unit weight error (RUWE), which measures the goodness of fit of the single-source astrometric model \citep{2021A&A...649A...5F}. These parameters are valuable indicators of problematic astrometric behaviour, but they do not constitute an independent measurement of optical morphology. They are coupled to the Gaia scanning law, windowing and image-parameter determination, calibration model, source brightness, and the astrometric solution itself \citep{2021A&A...649A...1G}. As a result, they cannot by themselves distinguish whether poor astrometric quality is caused by intrinsic source morphology, by observational effects, or by the Gaia processing pipeline. An independent morphological characterisation based on external optical imaging is therefore required \citep{2025arXiv251102204L}. Such a characterisation makes it possible to quantify how closely each Gaia-CRF3 source resembles an isolated point source, and to test directly whether the amplitude and form of Gaia astrometric systematics depend on optical morphology. This approach differs from selections based solely on Gaia astrometric quality indicators: instead of identifying problematic sources only after their astrometric solution has degraded, it introduces an external, image-based description of the source structure that can be used to assess the astrometric suitability of AGNs on a physical basis. This is essential for constructing cleaner reference-frame subsamples, for interpreting morphology-dependent parallax and proper motion systematics, and for guiding the source selection of future optical celestial reference frames.

% Following our previous work, we consider morphology to be a major factor governing the astrometric quality of AGNs. Multi-nucleus systems, AGNs with prominent host galaxies, and sources strongly contaminated by foreground stars all exhibit optical flux distributions that depart markedly from that of an ideal symmetric point source, and are therefore expected to compromise positional accuracy \citep{2017MNRAS.471.3775P, 2020ApJ...888...73H}. Gaia provides several parameters that are sensitive to such effects, including ipd\_frac\_multi\_peak, which reflects the probability that a source contains multiple components \citep{2022NatAs...6.1185M}, astrometric\_excess\_noise as an indicator of photocentre jitter or model mismatch \citep{2020ApJ...888...73H}, and the renormalised unit weight error (RUWE), which quantifies the residuals of the astrometric solution \citep{2021A&A...649A...5F}. These quantities, however, trace morphology only indirectly and are inevitably entangled with the Gaia observing process, as well as with the modelling assumptions and calibration prescriptions adopted in the data-processing pipeline \citep{2021A&A...649A...1G}. A more direct approach is therefore to characterise source morphology using imaging data from external optical surveys, and to use this information to assess astrometry stability in Gaia.

This study has two main objectives. First, we provide a physically motivated, imaging-based characterisation of the optical morphology of Gaia-CRF3 sources, complementing Gaia-internal quality indicators. Second, we quantify the impact of optical morphology on the astrometric stability of AGNs, with implications for the construction of cleaner optical reference-frame samples, the interpretation of parallax and proper motion systematics, and the source selection of future Gaia releases and wide-field astrometric surveys. The paper is organised as follows. Section \ref{sec_morphology} describes the construction of the point-source score and its validation, Section~\ref{sec_application} examines the dependence of Gaia-CRF3 astrometric quality and systematics on this score, and Section \ref{sec_conclusion} summarizes our conclusions.

%%%%%%%%%%%%%%%%%%%%%%%%%%%%%%%%%%%%%%%%%%%%%%%%%%%%%%%%%%%%%%

\section{Morphological Scores of AGNs}
\label{sec_morphology}
Many studies have shown that the abnormal optical morphology of AGNs can affect their astrometric stability, producing excess positional errors as well as spurious parallaxes or proper motions \citep{2020ApJ...888...73H, 2022ApJ...933...28M}. This is physically expected, because extended or asymmetric optical emission can displace the photocentre away from a stable compact nucleus and is generally less well represented by the point source image model used in Gaia image-parameter determination, thereby degrading centroiding precision and repeatability \citep{2016A&A...595A...3F,2012A&A...538A.107P,2020ApJ...888...73H}. This property has been widely exploited to select objects of interest, such as dual AGNs and lensed quasars, through simple parallax and proper motion cuts \citep{2019MNRAS.483.4242L, 2023MNRAS.524.1909J}. For assessing the stability of reference-frame sources, however, such empirical thresholds cannot fully eliminate morphologically disturbed objects. What is needed instead is a continuous, Gaia-independent morphological indicator that quantifies the point-source likeness of each object and enables a source-by-source assessment of its astrometric suitability.

The ideal morphology is that of an unresolved point source, whose observed profile should be fully described by the local point-spread function. The PS1-PSC provides point source probabilities for billions of sources \citep{2018PASP..130l8001T, 2021PASP..133e4502M}. By combining multiple morphological features through a machine-learning framework, such as \texttt{psfKronRatio} and \texttt{psfLikelihood}, it assigns each source a score between 0 and 1, where values closer to 1 indicate a higher probability of being morphologically point-like. In constructing this score, \citet{2018PASP..130l8001T} combined PS1 flux and shape measurements across the $grizy$ bands into signal-to-noise-weighted white-flux features, rather than relying on a single-band morphology. This multi-band treatment is well suited to point-source classification, because unresolved sources are expected to remain PSF-like across optical bands after band-dependent seeing, PSF, and noise effects are accounted for. Owing to its high reliability, this catalogue provides a reliable starting point for our analysis. Since Gaia currently offers no publicly available imaging data or morphological parameters, and host-galaxy detections are available for only a very small subset of sources, we constructed an independent all-sky morphological catalogue from Pan-STARRS DR1 (PS1) \citep{2002SPIE.4836..154K}, SDSS \citep{2000AJ....120.1579Y}, DESI Legacy Imaging Surveys \citep{2019AJ....157..168D}, and SkyMapper \citep{2018PASA...35...10W} to evaluate the morphology of Gaia-CRF3 sources. Using the PS1-PSC point-source score as the reference, we carefully selected the morphological parameters available in SDSS, DESI, and SkyMapper, and employed XGBoost to infer PSC-like scores for sources in these three surveys that are matched to Gaia-CRF3, while maintaining good consistency with PS1-PSC. In total, we provide morphological scores for 1,607,490 sources, whereas the remaining 6,683 sources are assigned null values because no usable morphological parameters are available.

\subsection{Morphological Score Construction for External Imaging Surveys}

The cross-match tables provided by the \href{https://gea.esac.esa.int/archive/}{Gaia Archive} contain source associations between Gaia DR3 and PS1, SDSS DR13, and SkyMapper DR2. Using the corresponding $best\_neighbor$ tables, we matched Gaia-CRF3 sources to these three catalogues. To ensure the use of the most recent SkyMapper release, we additionally performed a positional match to SkyMapper DR4 sources within a radius of 3 arcsec. For DESI, we relied on the precomputed Gaia-DESI cross-match table released by \href{https://noirlab.edu/public/}{NOIRLab}, $gaia\_dr3.x1p5\_gaia\_source\_ls\_dr10\_tractor$, to associate Gaia-CRF3 sources with DESI counterparts and thereby improve the reliability of the source matching.

XGBoost, or Extreme Gradient Boosting, is an efficient gradient-boosted decision-tree algorithm \citep{2016arXiv160302754C}. It sequentially trains a large number of shallow decision trees, with each subsequent tree fitted to the residuals of the previous stage, thereby combining multiple weak learners into a powerful predictive model. For each external imaging survey, the input morphological parameters were selected using a forward stepwise feature-selection procedure, with the validation-set root mean square error (RMSE) as the retention criterion. The detailed workflow is described in Appendix~\ref{sec:workflow}. This procedure preserves the consistency and comparability of the morphological scores across multiple data sources. The details for each catalogue are described below.

% Based on the matched morphological data, we selected, for each catalogue, a set of morphological parameters\footnote{We adopted a forward stepwise feature-selection procedure. Specifically, the candidate parameters were first ranked according to their correlation with the PS1-PSC score. Model training was then initialized with the single parameter showing the strongest correlation, and the prediction error was evaluated on the validation set. At each subsequent iteration, one previously unused parameter was added to the current feature set by selecting the candidate that produced the largest reduction in the validation-set error. The feature-selection procedure was terminated once the reduction in validation-set error brought by an additional parameter fell below 0.0001.} showing the strongest correlation with the PS1-PSC score, and used XGBoost to map the morphological parameters from SDSS, DESI, and SkyMapper onto the same scale as the PS1-PSC score. 

\subsubsection{SDSS}
\label{sec_sdss}
Based on the Gaia DR3 cross-match results, we extracted morphological parameters for 795,676 Gaia-CRF3 sources from the SDSS survey. Among them, 774,633 sources also have PS1-PSC scores, and these common sources form the training basis for the SDSS scoring model. During the XGBoost training process, we first stratified the sample by PS1-PSC score and then randomly split each stratum into the training, validation, and test sets with proportions of 0.70, 0.15, and 0.15, respectively. This strategy preserves the score distribution while improving the generalization capability of the final model.

The input features used for model training were selected as Appendix~\ref{sec:workflow}. This procedure led to the final selection of thirteen input parameters, as listed in Table \ref{tab:feature_physical_meaning}. Based on this, we trained an XGBoost regression model to map the SDSS morphological parameters onto the PS1-PSC score scale, with the optimal hyperparameters summarized in Table \ref{tab:sdsshyperparameters}. The model achieved an RMSE of 0.1423 on the validation set, while the RMSE difference between the training and validation sets was only 0.0063, indicating a stable fitting process with no clear sign of overfitting. On the test set, the model reached an RMSE of 0.1414 and a Mean Absolute Error (MAE) of 0.0760, and the Pearson correlation coefficient with the original PS1-PSC scores was 0.8334. These results show that the model not only provides accurate regression of the point-source score, but also preserves the relative ordering of source morphology on a continuous scale.

\begin{table}
\caption{SDSS input feature descriptions.}
\label{tab:feature_physical_meaning}
\centering
\small
\begin{tabular}{p{0.42\columnwidth} p{0.50\columnwidth}}
\hline\hline
\textbf{Feature} & \textbf{Description} \\
\hline
\texttt{is} & Point-source flag in the $i$ band. \\
\texttt{rdVrad} & Scale parameter of the de Vaucouleurs fit in the $r$ band. \\
\texttt{i\_psf\_minus\_petro} & PSF minus Petrosian magnitude in the $i$ band; traces source concentration. \\
\texttt{rdVell} & Ellipticity parameter of the de Vaucouleurs fit in the $r$ band. \\
\texttt{idVrad} & Scale parameter of the de Vaucouleurs fit in the $i$ band. \\
\texttt{z\_psf\_minus\_petro} & PSF minus Petrosian magnitude in the $z$ band; traces source concentration. \\
\texttt{rPrad} & Scale parameter of the exponential fit in the $r$ band. \\
\texttt{g\_psf\_minus\_petro} & PSF minus Petrosian magnitude in the $g$ band; traces source concentration. \\
\texttt{r\_psf\_minus\_petro} & PSF minus Petrosian magnitude in the $r$ band; traces source concentration. \\
\texttt{idVell} & Ellipticity parameter of the de Vaucouleurs fit in the $i$ band. \\
\texttt{zdVrad} & Scale parameter of the de Vaucouleurs fit in the $z$ band. \\
\texttt{gdVrad} & Scale parameter of the de Vaucouleurs fit in the $g$ band. \\
\texttt{iPrad} & Scale parameter of the exponential fit in the $i$ band. \\
\hline
\end{tabular}
\end{table}

\begin{table}
\caption{Optimized hyperparameters of the XGBoost model.}
\label{tab:sdsshyperparameters}
\centering
\footnotesize
\begin{tabular}{lccc}
\hline\hline
Parameter & SDSS & DESI & SkyMapper \\
\hline
\texttt{eta}                & 0.036 & 0.030& 0.045\\
\texttt{max\_depth}         & 5& 6& 4 \\
\texttt{min\_child\_weight} & 20.0 & 16.0& 24.0\\
\texttt{subsample}          & 0.70 & 0.90& 0.80\\
\texttt{colsample\_bytree}  & 0.70 & 0.80& 0.85\\
\texttt{alpha}              & 0.2 & 0.1& 0.2\\
\texttt{lambda}             & 1.0 & 1.5& 2.5\\
\texttt{gamma}              & 0.0 & 0.1& 0.0\\
\hline
\end{tabular}
\end{table}

\begin{figure}[ht!]
\centering
\includegraphics[width=\hsize]{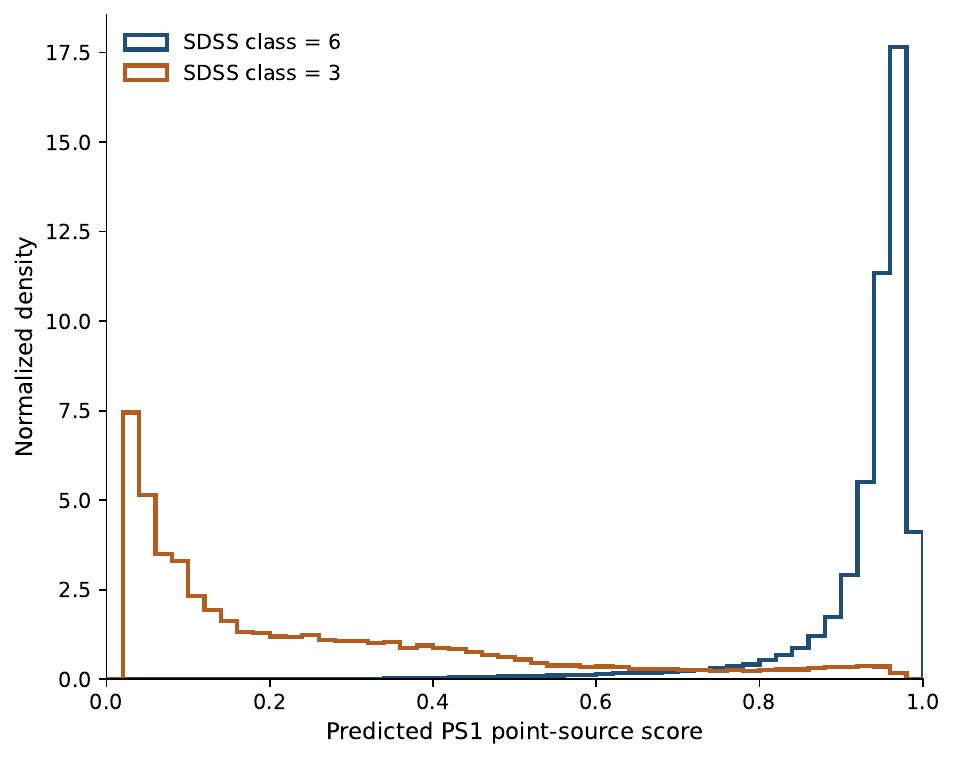}
    \caption{normalised density score distributions by SDSS morphological class. The area under each histogram is normalised to unity. }
     \label{fig.sdss_reliability}
\end{figure}

To examine the internal consistency of the point scores within the SDSS classification system, we further performed an internal validation using the SDSS labels, where \texttt{class} = 6 denotes point sources and \texttt{class} = 3 denotes extended sources, as shown in Fig. \ref{fig.sdss_reliability}. For the full sample, the model achieved an Area Under the Curve (AUC) of 0.9857, with a mean score of 0.2409 for extended sources and 0.9274 for point sources. In the higher-quality subsample with clean = 1, the AUC further increased to 0.9868, while the other main statistics remained stable, indicating that the results are not driven by a small number of anomalous observations. The precision and recall for point-source identification under different score thresholds are listed in Table \ref{tab:sdss_ps_metrics}. Taken together, the independent test results and the internal consistency check demonstrate that the SDSS-based morphological score is highly reliable and can serve as an important component of a unified morphological scoring framework across multiple surveys.

\begin{table}
\caption{Point-source precision and recall for different SDSS samples and point score thresholds.}
\label{tab:sdss_ps_metrics}
\centering
\small
\begin{tabular}{lccc}
\hline\hline
Sample & Threshold & Precision & Recall \\
\hline
All samples & 0.5 & 0.9883 & 0.9911 \\
All samples & 0.8 & 0.9953 & 0.9280 \\
\texttt{clean=1} subsample     & 0.5 & 0.9892 & 0.9912 \\
\texttt{clean=1} subsample     & 0.8 & 0.9957 & 0.9291 \\
\hline
\end{tabular}
\end{table}

\subsubsection{DESI}
As in Sec. \ref{sec_sdss}, we obtained the morphological parameters of 1,426,260 Gaia-CRF3 sources from the DESI Legacy Surveys DR10 Tractor catalogue through the NOIRLab Astro Data Lab. Among them, 1,122,446 sources also have PS1-PSC scores, and these common sources form the training basis for the DESI scoring model. We followed the same procedure as in Sec. \ref{sec_sdss}, including the construction of the training, validation, and test sets, the selection of input features, and the optimization of the XGBoost hyperparameters, with the detailed results listed in Tables \ref{tab:desifeature_physical_meaning} and \ref{tab:sdsshyperparameters}. The model achieved an RMSE of 0.1397 on the validation set, while the RMSE difference between the training and validation sets was 0.0031, indicating a stable fitting process. Compared with SDSS, DESI yields better PS1-PSC regression performance on the independent test set, with a test-set RMSE of 0.1378, an MAE of 0.0743, and a Pearson correlation coefficient of 0.8521. We applied the model to all 1,426,260 DESI sources and generated the corresponding scoring table.

\begin{table}
\caption{DESI input feature descriptions.}
\label{tab:desifeature_physical_meaning}
\centering
\small
\begin{tabular}{p{0.42\columnwidth} p{0.50\columnwidth}}
\hline\hline
\textbf{Feature} & \textbf{Description} \\
\hline
\texttt{tractor\_is\_psf} & Binary indicator that the Tractor solution prefers the PSF model \\
\texttt{shape\_r} & Morphological scale radius in the $r$ band given by the Tractor model \\
\texttt{log1p\_model\_margin} & Log-transformed margin between the best-fit model and the runner-up model \\
\texttt{log1p\_snr\_median} & Log-transformed median signal-to-noise ratio of the photometric measurements \\
\texttt{fracflux\_max} & Maximum fractional flux contribution measured\\
\texttt{fitbits} & Bitwise quality flags that summarize fit conditions and catalogue warnings \\
\texttt{log1p\_model\_preference} & Log-transformed preference of the best extended model over the PSF model \\
\texttt{log1p\_rchisq\_excess} & Reduced-$\chi^2$ excess above unity\\
\texttt{fracin\_deficit} & Deficit of the in-aperture flux fraction, tracing extended or off-center light profiles \\
\texttt{fracmasked\_max} & Maximum fraction of masked pixels\\
\hline
\end{tabular}
\end{table}

The internal consistency test further demonstrates that the score is highly self-consistent within the DESI system. When grouped by DESI tractor\_type, the samples classified as Point Sources (PSF) reach a median score of 0.9579, whereas the extended-source models remain substantially lower, with median scores of 0.0568 for Sersic profiles (SER), 0.2987 for round exponential galaxies with a variable radius (REX), and 0.1960 for DeVaucouleurs profiles (DEV), indicating a clear structural separation. If tractor\_type = PSF is treated as a label for point sources and all remaining types are merged into the non-point-source class, the internally derived AUC based on the full predicted score table reaches 0.9972. The precision and recall for point-source identification under different score thresholds are listed in Table \ref{tab:desi_ps_metrics}. Although this test still represents an internal consistency check within the survey, the regression performance on the independent test set indicates that the DESI-based morphological score is highly reliable, and that its overall performance is slightly better than that of the SDSS-based score.

\begin{table}
\caption{Point-source precision and recall for different DESI samples and point score thresholds.}
\label{tab:desi_ps_metrics}
\centering

\begin{tabular}{ccc}
\hline\hline
Threshold & Precision & Recall \\
\hline
0.5 & 0.9889 & 0.9954 \\
0.8 & 0.9997 & 0.9212 \\
0.9 & 1.0000 & 0.8111 \\
\hline
\end{tabular}
\end{table}

\subsubsection{SkyMapper}

SkyMapper provides an important morphological supplement for the southern sky. We matched a total of 763,195 Gaia-CRF3 sources to this survey, among which 458,171 also have PS1-PSC scores. Unlike SDSS and DESI, however, the morphological parameters in SkyMapper are substantially less complete. We therefore ranked the candidate SkyMapper parameters by their correlation with the PS1-PSC score and adopted a forward stepwise feature-selection procedure. This procedure retained the eleven parameters listed in Table \ref{tab:skymapperfeature_physical_meaning}. Among the matched SkyMapper sources, 234,842 objects, corresponding to 30.8\%, have complete measurements for all eleven parameters. The remaining sources were still usable because XGBoost handled the missing values directly during training and prediction.

\begin{table}
\caption{SkyMapper input feature descriptions.}
\label{tab:skymapperfeature_physical_meaning}
\centering
\small
\begin{tabular}{p{0.45\columnwidth} p{0.45\columnwidth}}
\hline\hline
\textbf{Feature} & \textbf{Description} \\
\hline
\texttt{z\_psf\_minus\_petro} & Difference between the PSF and Petrosian magnitudes in the \textit{z} band \\
\texttt{gri\_psf\_minus\_petro\_median} & Median PSF minus Petrosian magnitude difference across the \textit{g}, \textit{r}, and \textit{i} bands \\
\texttt{one\_minus\_classstar} & Complement of the point-source classification metric \texttt{ClassStar} \\
\texttt{log1p\_chi2PSF} & Log-scaled quantity of the PSF fitting goodness \\
\texttt{optical\_mag\_median} & Median PSF magnitude across the \textit{g}, \textit{r}, and \textit{i} bands \\
\texttt{FWHM} & Full width at half maximum of the source image \\
\texttt{flags\_max\_griz} & Maximum quality flag across the \textit{g}, \textit{r}, \textit{i}, and \textit{z} bands \\
\texttt{i\_psf\_minus\_petro} & Difference between the PSF and Petrosian magnitudes in the \textit{i} band \\
\texttt{r\_psf\_minus\_petro} & Difference between the PSF and Petrosian magnitudes in the \textit{r} band \\
\texttt{RadrPetro} & Petrosian radius of the source \\
\texttt{log1p\_Ngood} & Log-scaled number of good detections \\
\hline
\end{tabular}
\end{table}

Because the seeing quality of SkyMapper is poorer than that of the other surveys considered here, and its morphological parameters are more incomplete, the regression performance is weaker than that obtained for SDSS and DESI. On the test set, the model yields an RMSE of 0.1857, an MAE of 0.0885, and a Pearson correlation coefficient of 0.6550. After applying the model to the full sample, we further examined the internal consistency of the SkyMapper-based scores. Specifically, we treated SkyMapper \texttt{ClassStar} $\le 0.2$ and \texttt{ClassStar} $\ge 0.95$ as internal extreme pseudo-labels for extended and point-like sources, respectively. For the subset of 181,690 sources with available ClassStar values, the resulting AUC is 0.9421. The mean score is 0.6749 for extended sources and 0.9533 for point-like sources. 

These results indicate that, although the regression accuracy of SkyMapper is lower than that of DESI and SDSS, its score still provides useful supplementary coverage, especially in the southern sky, albeit with larger calibration uncertainty. It is therefore used as an auxiliary source of morphological information when PS1, SDSS, or DESI based scores are unavailable for a given source.

\subsection{Cross-Survey Integration and Reliability Assessment}
\label{cross_survey}
\begin{figure}[ht!]
\centering
\includegraphics[width=\hsize]{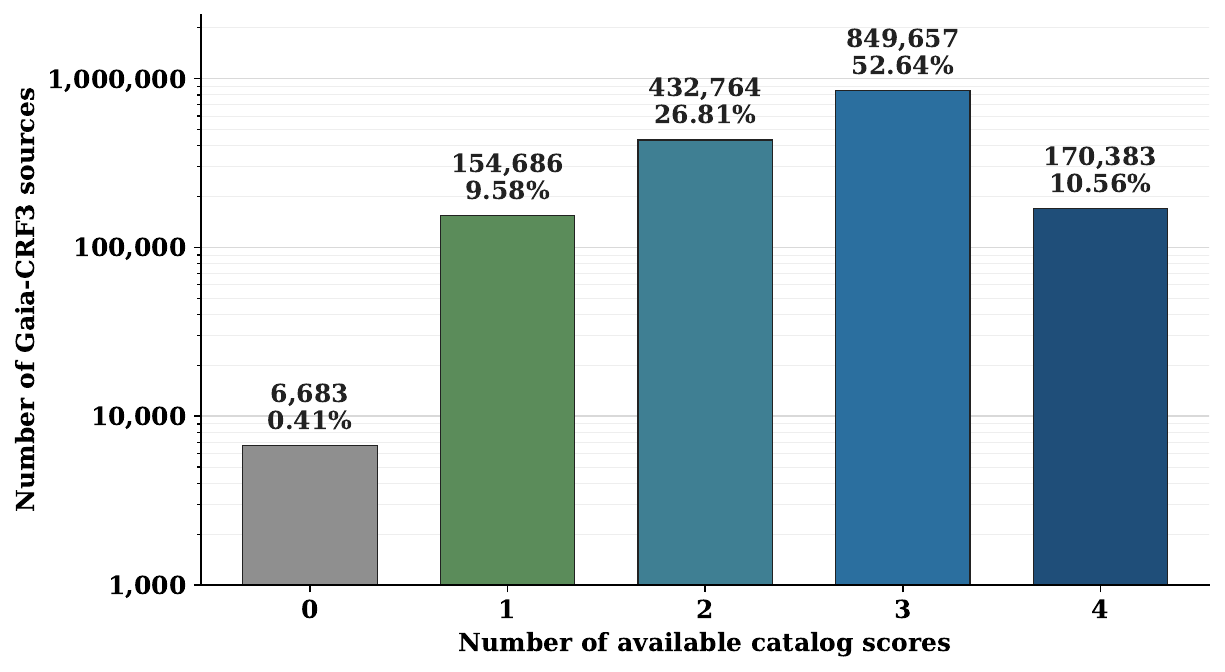}
    \caption{Distribution of Gaia-CRF3 sources by the number of available catalogue-based morphology scores. Bar labels give the number and fraction of sources in each category.
}
     \label{fig.number}
\end{figure}

\begin{figure}[ht!]
\centering
\includegraphics[width=\hsize]{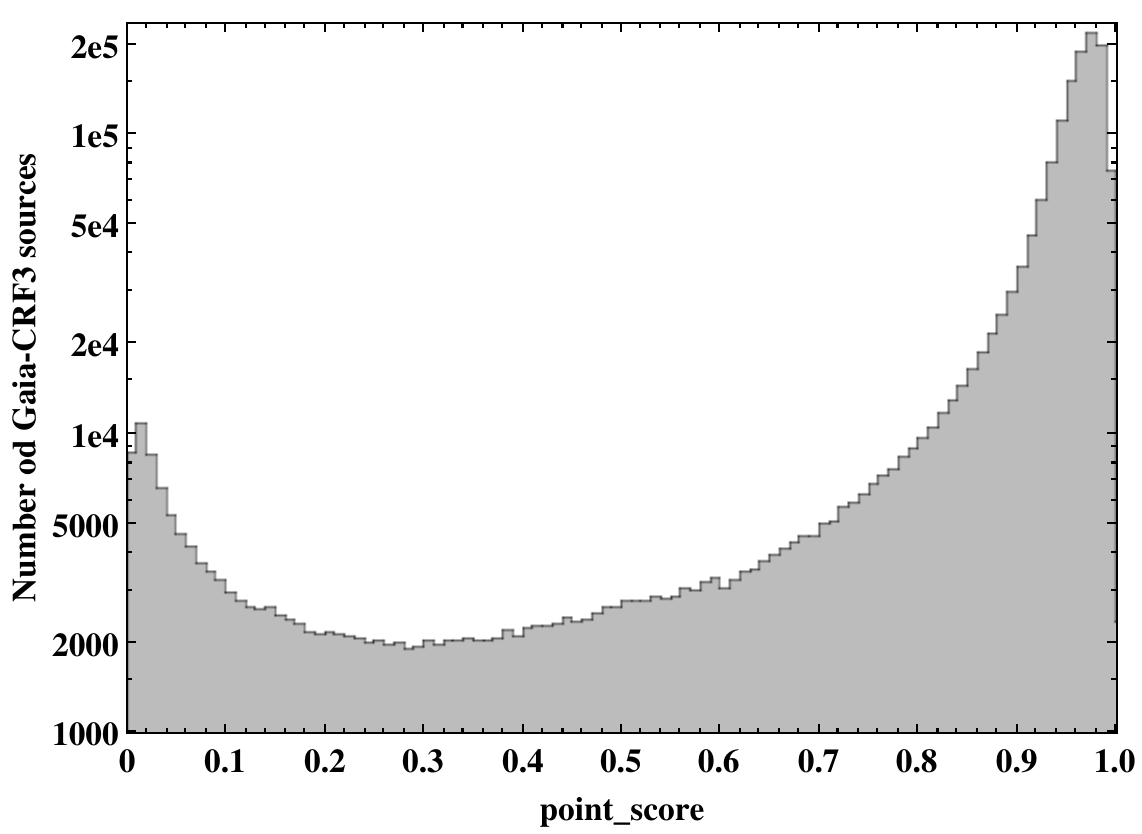}
    \caption{Distribution of the \texttt{point\_score} of Gaia-CRF3 sources.
}
     \label{fig.distribution}
\end{figure}

By cross-matching Gaia-CRF3 with external survey catalogues and performing regression with the XGBoost models, we ultimately obtained a point-source score from at least one external catalogue for 1,607,490 Gaia-CRF3 sources, corresponding to a completeness of 99.59\% with respect to the full Gaia-CRF3 catalogue. The coverage of morphology-score information for the Gaia-CRF3 sources is shown in Fig. \ref{fig.number}. For ease of use, we retained the individual point-source scores from the different surveys and also provided a composite point-source score (\texttt{point\_score} hereafter) for each Gaia-CRF3 source. Because the reliability of point-source discrimination depends strongly on the seeing conditions and limiting magnitude of each survey, we initially adopted the priority order DESI, PS1, and SDSS when assigning the composite score, while a SkyMapper-based score was used only when no valid score was available from the other three surveys. However, through visual inspection, we found that a fixed survey-priority scheme can amplify anomalous measurements from an individual survey. In a small number of cases, DESI assigns relatively high scores to sources that are likely extended, whereas PS1, SDSS, and SkyMapper consistently return much lower values, indicating that cross-survey disagreement itself carries useful diagnostic information. We therefore adopted a more robust multi-survey fusion strategy. In this revised scheme, DESI remains the most influential source of morphological information, but it no longer has absolute priority. When valid scores are available from multiple surveys, the final score is determined jointly from all available measurements, while strongly conflicting cases are down-weighted or flagged. For the small subset of sources in which DESI gives a high score but the other surveys consistently support a much lower value, we adopt a more conservative combined score in order to reduce the impact of misclassification from any single survey on the final morphological assessment. The full definition of the fusion strategy and its implementation are provided in Appendix~\ref{sec:fusion}. The final \texttt{point\_score} distribution is shown in Fig. \ref{fig.distribution}.

\begin{figure}[ht!]
\centering
\includegraphics[width=\hsize]{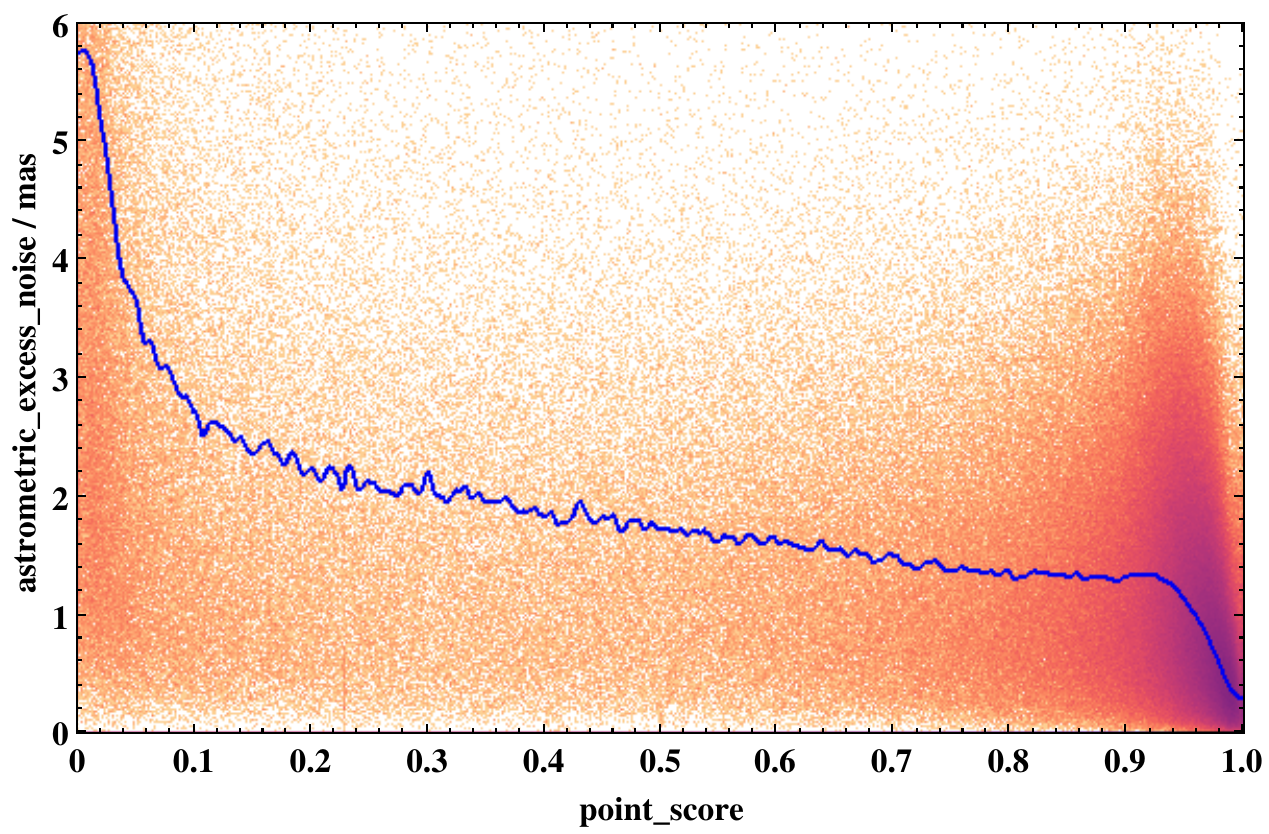}
    \caption{Distribution of \texttt{astrometric\_excess\_noise} versus \texttt{point\_score} for Gaia-CRF3 sources. Each point represents one source, and darker colours indicate higher source density. The blue curve shows the 75th percentile in each \texttt{point\_score} bin. Since 55.46\% of the sources with \texttt{point\_score} $> 0.94$ have zero \texttt{astrometric\_excess\_noise}, a median curve would become identically zero in this regime and would therefore fail to reflect the underlying trend.
}
     \label{fig.excess}
\end{figure}

\begin{figure}[ht!]
\centering
\includegraphics[width=\hsize]{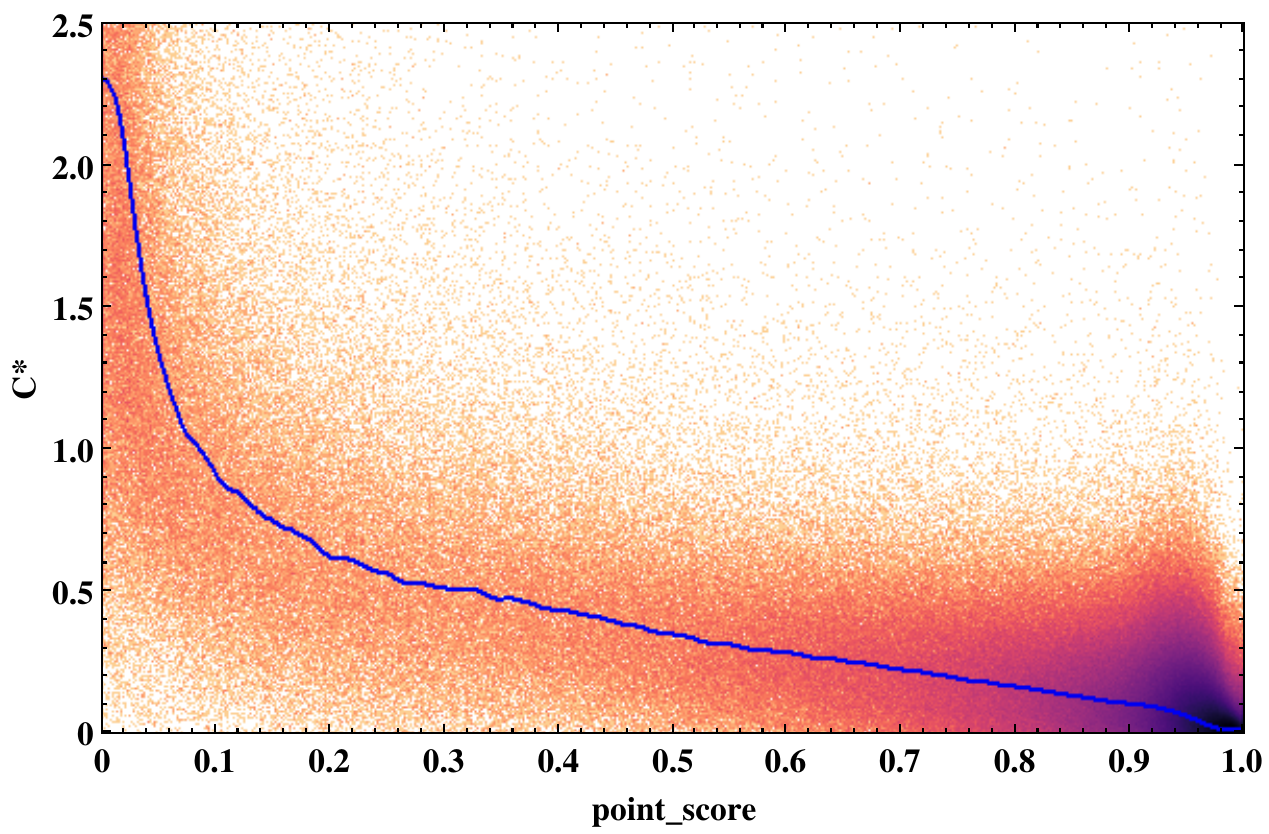}
    \caption{Distribution of the corrected BP/RP flux-excess factor, $C^*$,  versus \texttt{point\_score} for Gaia-CRF3 sources. Each point represents one source, and darker colours indicate higher source density. The blue curve shows the median value in each \texttt{point\_score} bin.
}
     \label{fig.cstar}
\end{figure}

In the previous sections, we evaluated the internal consistency of the DESI, SDSS, and SkyMapper point-source scores, as well as their consistency with the PS1-PSC score, and showed that these scores preserve the same descriptive scale as PS1-PSC while maintaining good internal coherence. Gaia also provides several parameters that can indirectly reflect the astrometric quality of a source. For example, \texttt{astrometric\_excess\_noise} absorbs contributions from calibration errors, part of the high-frequency attitude noise, and residuals caused by deviations of the source morphology from the single-star model. Likewise, the colour-corrected \texttt{phot\_bp\_rp\_excess\_factor} ($C^*$) reflects effects related to background-estimation residuals, contamination from nearby sources, and blending in crowded fields. These quantities are all expected to be sensitive to departures from a point-like morphology. We therefore examined the variation of \texttt{point\_score} with respect to these two Gaia parameters, as shown in Figs. \ref{fig.excess} and \ref{fig.cstar}. The point-source score exhibits a clear negative correlation with both quantities, further indicating that \texttt{point\_score} is an effective tracer of the deviation of a source morphology from that of an ideal point source, and that it is highly reliable.

\begin{figure}[ht!]
\centering
\includegraphics[width=\hsize]{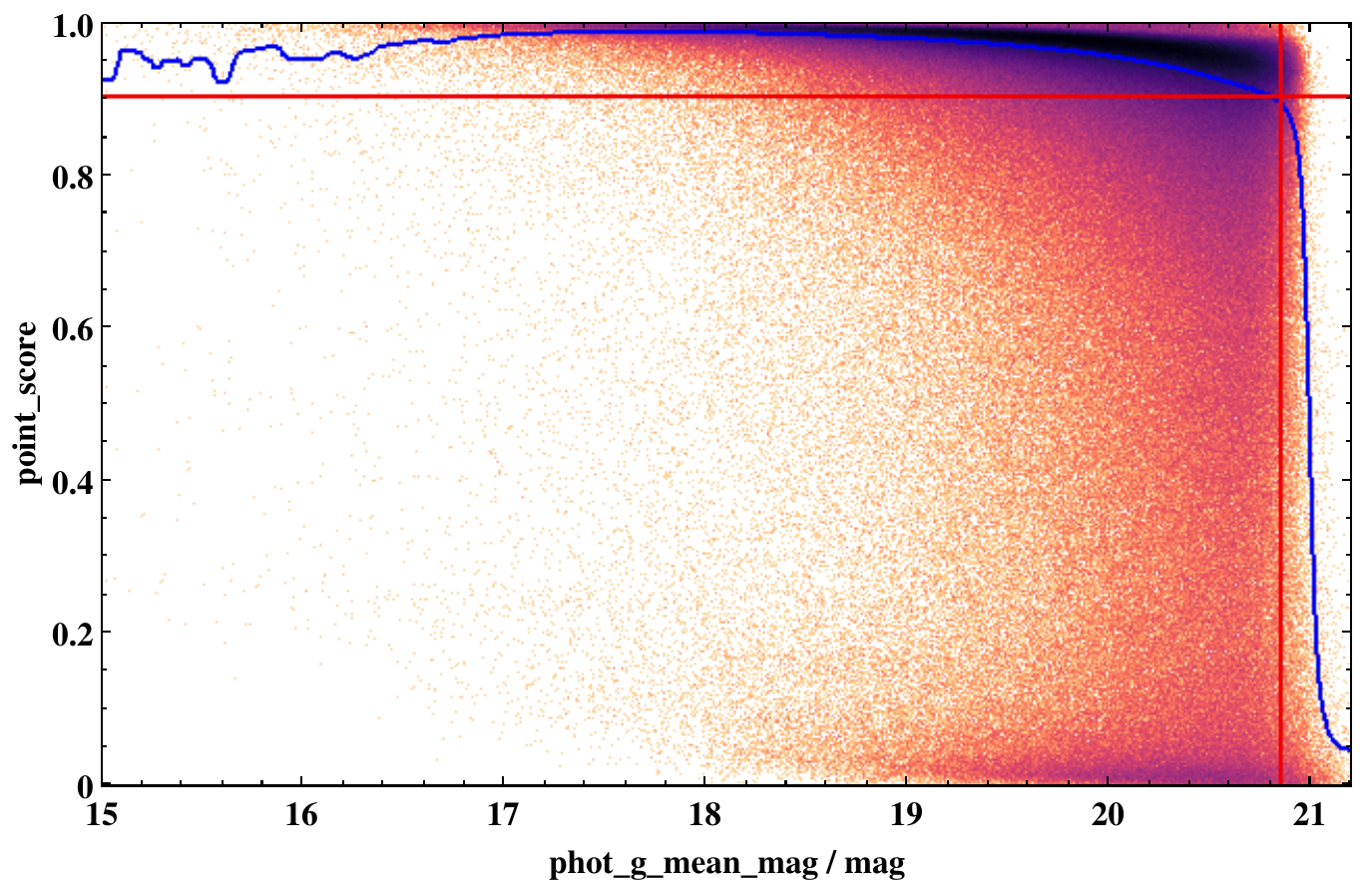}
    \caption{Distribution of Gaia $G$ magnitude versus \texttt{point\_score} for Gaia-CRF3 sources. Each point represents one source, and darker colours indicate higher source density. The blue curve shows the median value in each magnitude bin. The two red lines mark \texttt{point\_score} $= 0.9$ and \texttt{phot\_g\_mean\_mag} $= 20.85$ mag, respectively.
}
     \label{fig.gamg_point}
\end{figure}

The morphological classification inferred from external survey images also depends strongly on image signal-to-noise ratio. This may cause faint point-like sources to be misclassified as extended objects and therefore assigned artificially low \texttt{point\_score} values. We therefore examined whether \texttt{point\_score} shows any systematic dependence on magnitude. Fig. \ref{fig.gamg_point} presents the distribution of \texttt{point\_score} as a function of Gaia $G$ magnitude, together with the trend of the median value. Although the median begins to decline beyond $G \sim 19$, it remains above 0.9 for $G < 20.85$, indicating that the score remains effective over this magnitude range. For $G > 20.85$, the median drops rapidly. This is partly because the number of Gaia sources becomes very small in this regime, and partly because the external-survey observations also have low signal-to-noise ratio. As a result, the \texttt{point\_score} is therefore not considered reliable in this faint regime.

\begin{figure*}[!t]
\centering
\includegraphics[width=\hsize]{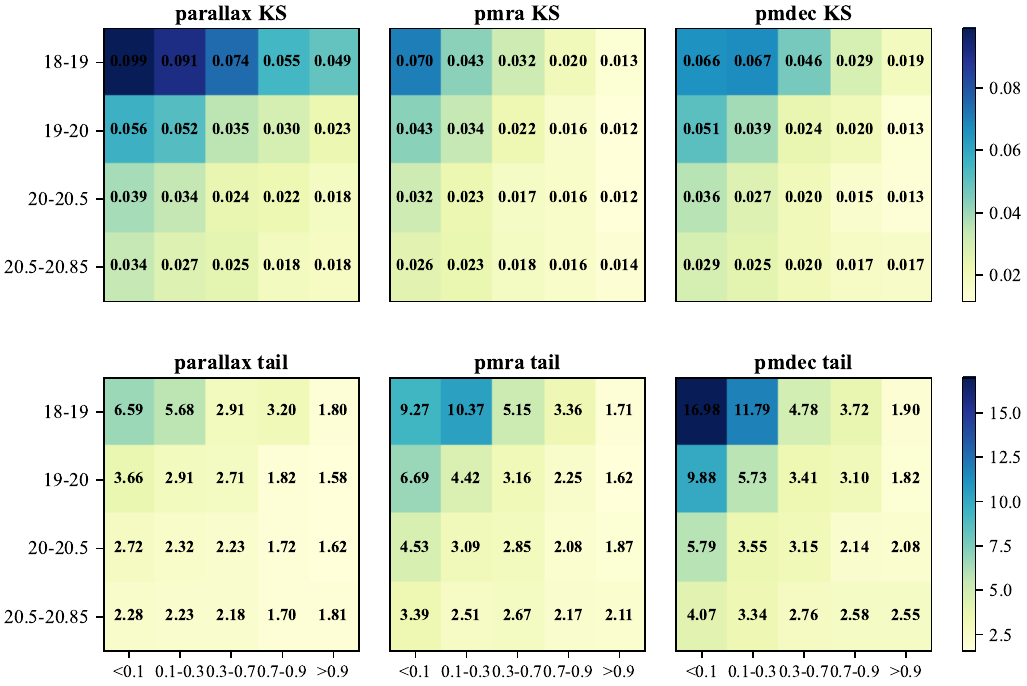}
    \caption{Heat maps of the deviations of the normalised Gaia astrometric quantities from a standard normal distribution as a function of point-score bin and $G$-magnitude bin. The horizontal axis gives the point-score bins, from $<0.1$ to $>0.9$, and the vertical axis gives the Gaia $G$-magnitude bins, from $18$--$19$ to $20.5$--$20.85$. The three columns show $\varpi/\sigma_\varpi$, $\mu_{\alpha *}/\sigma_{\mu_{\alpha *}}$, and $\mu_\delta/\sigma_{\mu_\delta}$, respectively. The top row gives the Kolmogorov--Smirnov distance between the empirical distribution and the standard normal distribution, while the bottom row shows the tail-excess factor. The numbers in each cell give the corresponding measured values.
}
     \label{fig.gauss_papm}
\end{figure*}

The primary purpose of assigning morphological scores to Gaia-CRF3 sources is to clarify the morphological properties of the current reference-frame sources and to identify reliable point-like objects that can help reduce systematic errors in the reference frame. An ideal reference source should have parallax and proper motion values close to zero, and its normalised parallax and normalised proper motion components should follow the standard normal distribution. To quantify the deviation of the normalised parallax and proper motion components from the standard Gaussian distribution $N(0,1)$, we adopted two statistical measures. First, the Kolmogorov--Smirnov distance \citep{massey1951kolmogorov} is defined as
\begin{equation}
D_{\rm KS} = \sup_x \left| F_n(x) - \Phi(x) \right| ,
\end{equation}
where $F_n(x)$ is the empirical cumulative distribution function of the sample and $\Phi(x)$ is the cumulative distribution function of the standard normal distribution. Smaller values of $D_{\rm KS}$ indicate that the sample distribution is globally closer to the standard Gaussian. Second, motivated by the excess non-Gaussian tails seen in the normalised astrometric parameters of Gaia-CRF3 sources \citep{2022A&A...667A.148G}, we define the tail-excess factor for $|z| > 3$ as
\begin{equation}
E_{\rm tail} = \frac{P(|z|>3)}{2\left[1-\Phi(3)\right]} ,
\end{equation}
where the denominator is the theoretical fraction of a standard Gaussian beyond $|z| > 3$, approximately 0.0027. Values of $E_{\rm tail} > 1$ indicate that the sample has heavier tails than the standard Gaussian and therefore contains more significant outliers.

Fig. \ref{fig.gauss_papm} shows the deviation of the normalised parallax and proper motion distributions from the Gaussian expectation in different magnitude bins, and the normalised histograms of the astrometric parameters are presented in Appendix \ref{sec:normalised}. The numerical values differ substantially from one magnitude range to another, mainly because the formal Gaia uncertainties vary strongly with source brightness. Nevertheless, within each magnitude bin, both $D_{\rm KS}$ and $E_{\rm tail}$ decrease markedly with increasing \texttt{point\_score}. This indicates that high-score sources consistently select subsamples with smaller systematic errors in every magnitude range, thereby providing a reliable basis for the construction of a high-precision celestial reference frame.

%%%%%%%%%%%%%%%%%%%%%%%%%%%%%%%%%%%%%%%%%%%%%%%%%%%%%%%%%%%%%%

\section{Applications of the Morphological Score}
\label{sec_application}

The morphological score derived in the previous section provides an objective and continuous external indicator for the full Gaia-CRF3 sample, characterising the intrinsic morphological properties of the sources while remaining independent of the Gaia instrument and data-processing pipeline. Such an external parameter provides an independent way to assess the impact of source morphology on the systematic errors of Gaia-CRF3, and thus to determine whether this effect needs to be explicitly taken into account in the construction of a higher-precision reference frame. In this section, we examine the influence of morphological properties on Gaia-CRF3 astrometry from the perspectives of both parallax and proper motion.

\subsection{Gaia Parallax Zero Point}
\label{sec:zero_point}
\begin{figure*}[!t]
\centering
\includegraphics[width=\hsize]{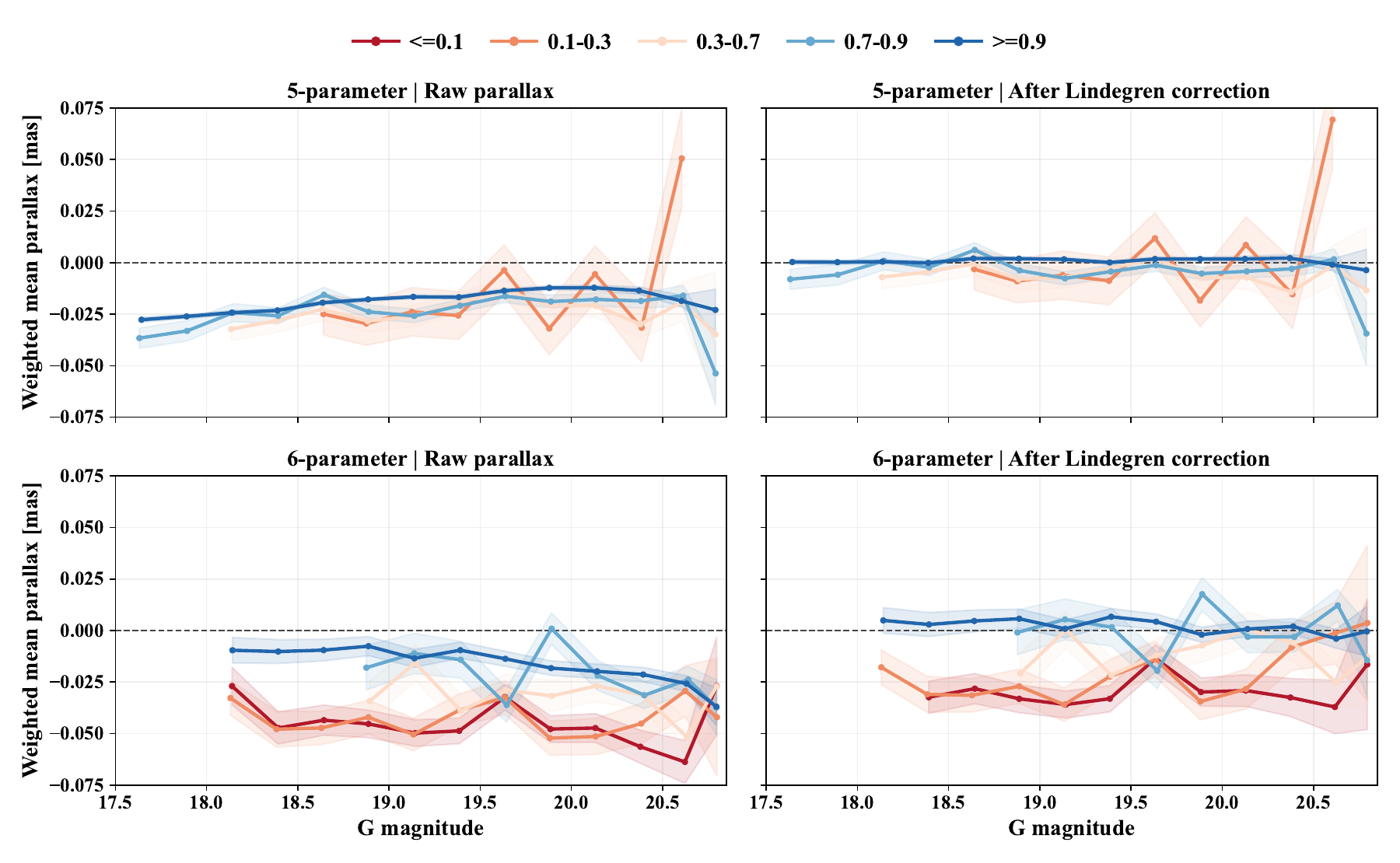}
    \caption{Weighted mean parallax as a function of $G$ magnitude for subsamples defined by different \texttt{point\_score}. The upper and lower panels correspond to the five-parameter and six-parameter solutions, respectively. The left column shows the raw parallaxes, while the right column shows the results after applying the parallax zero-point correction from \citet{2021A&A...649A...4L}. The weighted mean was computed using weights of $1/\sigma_\varpi^2$ over the magnitude range $17.5 \leq G \leq 20.85$. Because the number of sources with \texttt{point\_score} $\leq 0.1$ is extremely small in the five-parameter sample, this interval is not shown in the corresponding panels.
}
     \label{fig.parallax_mean}
\end{figure*}

Both the determination of the Gaia parallax zero point and the construction of its correction model \citep{2021A&A...649A...4L, 2024A&A...691A..81D} rely on source samples whose true parallaxes are expected to be close to zero. The purity of the point-source sample therefore directly affects the stability of the zero-point estimate and the robustness of the correction. When the Gaia-CRF3 sample is grouped by \texttt{point\_score}, the weighted mean parallax exhibits a clear monotonic stratification, see Fig. \ref{fig.parallax_mean}. In the range $18 \leq G < 20.5$, the weighted mean of the raw parallaxes changes from $-43.7\,\mu$as for sources with \texttt{point\_score} $\leq 0.1$ to $-17.5\,\mu$as for sources with \texttt{point\_score} $\geq 0.9$, corresponding to a difference of 26.2~$\mu$as. After applying the parallax zero-point correction based on magnitude, ecliptic latitude, and colour, the high-score sample moves back to nearly zero, with a weighted mean of $1.3\,\mu$as, whereas the low-score sample remains at $-26.9\,\mu$as, retaining a negative residual of 28.2~$\mu$as relative to the high-score group. This result demonstrates a stable connection between \texttt{point\_score} and the parallax zero-point residual. It shows that \texttt{point\_score} can identify cleaner samples that are better suited for zero-point anchoring, while also revealing morphology-related systematics that are not yet explicitly captured by the current zero-point model. At magnitudes fainter than $G=20.5$, especially at the faint end of the six-parameter sample, the mean parallaxes of different \texttt{point\_score} subsamples become nearly indistinguishable, and the systematic offset of the \texttt{point\_score} $\leq 0.1$ sample is even slightly smaller than that of the \texttt{point\_score} $\geq 0.9$ sample. After applying the parallax zero-point correction \citep{2021A&A...649A...4L}, however, the mean parallax of the high-score sample is brought close to zero, whereas the low-score sample still retains a residual offset of about $-20\,\mu$as. This indicates that, at the faint end, the dominant parallax systematics are governed by magnitude, colour, ecliptic latitude, and the Gaia astrometric solution itself, while the contribution from morphology is no longer significant.

Fig. \ref{fig.parallax_mean} shows that sources with \texttt{point\_score} $< 0.1$ are almost absent from the five-parameter sample, a pattern that is fully consistent with the Gaia astrometric processing scheme. The Gaia DR3 five-parameter solution is constructed under the condition that the colour information is sufficiently reliable, in which case the chromaticity term has already been handled at the image-location stage. By contrast, the six-parameter solution is used for sources with insufficient or unreliable colour information, and therefore requires the additional estimation of \texttt{pseudocolour} as a sixth astrometric parameter. The Gaia DR3 validation documentation further notes that six-parameter solutions occur more frequently in crowded regions and in areas with poorer early scanning coverage, while the standard Gaia astrometric model assumes that all sources are isolated point sources. Once this assumption is no longer well satisfied, the fit quality degrades and sources become more likely to receive a six-parameter solution. In the present sample, low-\texttt{point\_score} sources are more likely to contain host-galaxy structure, extended morphology, or contamination from nearby objects, and they therefore show a strong statistical preference for the six-parameter solution. In the subsample with $18 \leq G < 20.5$ and \texttt{point\_score} $\leq 0.1$, the fraction of six-parameter solutions reaches 96.7\%, which is a direct manifestation of the processing mechanism described above.

These results further show that \texttt{point\_score} is valuable in two closely related ways. On the one hand, it measures the degree to which a source departs from the point-source conditions for which Gaia astrometry performs best. On the other hand, it is directly linked to the parallax zero-point residual. \texttt{point\_score} can therefore be used both to prioritize high-purity sources for zero-point anchoring and as an additional variable beyond the current zero-point model to characterise the more prominent morphology-related residuals in the six-parameter sample.

\subsection{Gaia Reference Frame Systematic Errors}

Figure \ref{fig.gauss_papm} indicates that the imprint of AGN morphology on Gaia astrometry is not confined to parallax, but is also present in the proper motion measurements. In Section \ref{sec:zero_point}, we showed that morphologically anomalous AGNs can produce a parallax zero-point shift of approximately $-43.7~\mu$as, substantially larger than the offset measured for nominally point-like sources. Motivated by this result, we now examine the corresponding morphology-dependent systematics in proper motion and assess their impact on the realization of the Gaia reference frame. Specifically, we first quantify the spurious proper motions associated with different AGN morphological classes, and then investigate how these signals project onto the global proper motion field and the inferred frame spin. Although the official Gaia-CRF3 spin solution is based on about 0.43 million high-precision five-parameter sources listed in the table \texttt{gaiaedr3.frame\_rotator\_source}, characterising the spin behaviour of the full Gaia-CRF3 sample remains essential. This is particularly relevant for studies that use large AGN samples to fit all-sky proper motion fields, such as the analysis presented by \citet{2025NatAs...9.1396M}.

\subsubsection{Morphology-Related Proper Motion Systematics Errors}

\begin{figure}[ht!]
\centering
\includegraphics[width=\hsize]{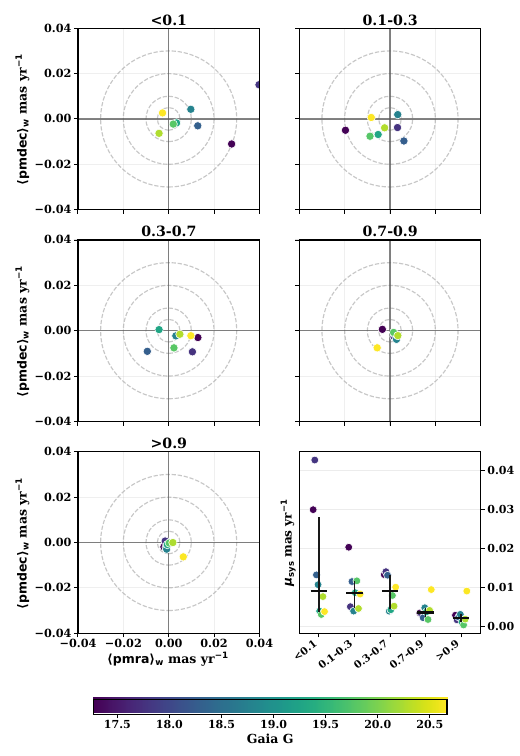}
    \caption{Weighted mean proper motion systematics in bins of the source morphology score \texttt{point\_score}. The first five panels show $\langle{\rm pmra}\rangle_{\rm w}$ and $\langle{\rm pmdec}\rangle_{\rm w}$ for the five score intervals, using the common Gaia $G$ magnitude bins from 17.0 to 20.85 mag. Points are colour-coded by the mean Gaia $G$ magnitude of each bin, and dashed circles mark fixed offsets from zero in mas yr$^{-1}$. The lower-right panel summarizes $\mu_{\rm sys} = (\langle{\rm pmra}\rangle_{\rm w}^2+\langle{\rm pmdec}\rangle_{\rm w}^2)^{1/2}$ for each score interval, black bars indicate the median and the 16th--84th percentile range across magnitude bins.}
     \label{fig.pm_systematics_grid}
\end{figure}

In this section, we quantify the proper motion systematics associated with source morphology. For Gaia-CRF3 sources with $17 \leq G < 20.85$, we divided the sample into eight equal-width magnitude bins and computed the weighted mean proper motions for different \texttt{point\_score} intervals in each bin. The full results are shown in Fig.~\ref{fig.pm_systematics_grid}. The first five panels of Fig.~\ref{fig.pm_systematics_grid} reveal a clear stratification of the residual proper motion systematics with \texttt{point\_score}, with the declination component showing a pronounced negative offset. Defining
\begin{equation}
\mu_{\rm sys} = \left( \langle {\rm pmra} \rangle_{\rm w}^{2} + \langle {\rm pmdec} \rangle_{\rm w}^{2} \right)^{1/2},
\end{equation}
the corresponding distribution is shown in the lower-right panel of Fig.~\ref{fig.pm_systematics_grid}. For the five \texttt{point\_score} intervals $<0.1$, $0.1$--$0.3$, $0.3$--$0.7$, $0.7$--$0.9$, and $>0.9$, the median values of $\mu_{\rm sys}$ are 9.1, 8.4, 9.0, 3.6, and 2.1~$\mu$as~yr$^{-1}$, respectively. The highest-score sample therefore shows a reduction of about 77\% relative to the lowest-score sample, indicating a strong connection between \texttt{point\_score} and the residual proper motion systematics, with high-score sources exhibiting substantially quieter systematic offsets.

After excluding the faintest bin, $20.5 < G < 20.85$~mag, this trend becomes more pronounced, with median $\mu_{\rm sys}$ values of 10.7, 8.6, 7.9, 3.4, and 1.9~$\mu$as~yr$^{-1}$ from the lowest- to the highest-score interval. The faintest bin behaves differently: its $\mu_{\rm sys}$ values are 3.8, 8.3, 10.0, 9.4, and 9.0~$\mu$as~yr$^{-1}$ across the same five score intervals, and thus do not decrease with increasing score. In particular, the \texttt{point\_score} $>0.9$ subsample still reaches $\mu_{\rm sys}=9.0$~$\mu$as~yr$^{-1}$, corresponding to $\langle {\rm pmra} \rangle_{\rm w}=6.5$~$\mu$as~yr$^{-1}$ and $\langle {\rm pmdec} \rangle_{\rm w}=-6.3$~$\mu$as~yr$^{-1}$. This suggests that, near the Gaia magnitude limit, measurement noise, sample-selection effects, or cross-match quality begin to dominate the residual systematics, substantially weakening the ranking power of the morphological score for proper motion systematics. This behaviour is consistent with the parallax zero-point systematics discussed in Section~\ref{sec:zero_point}.

\subsubsection{Frame Spin and the Proper Motion Field}
\begin{figure}[ht!]
\centering
\includegraphics[width=\hsize]{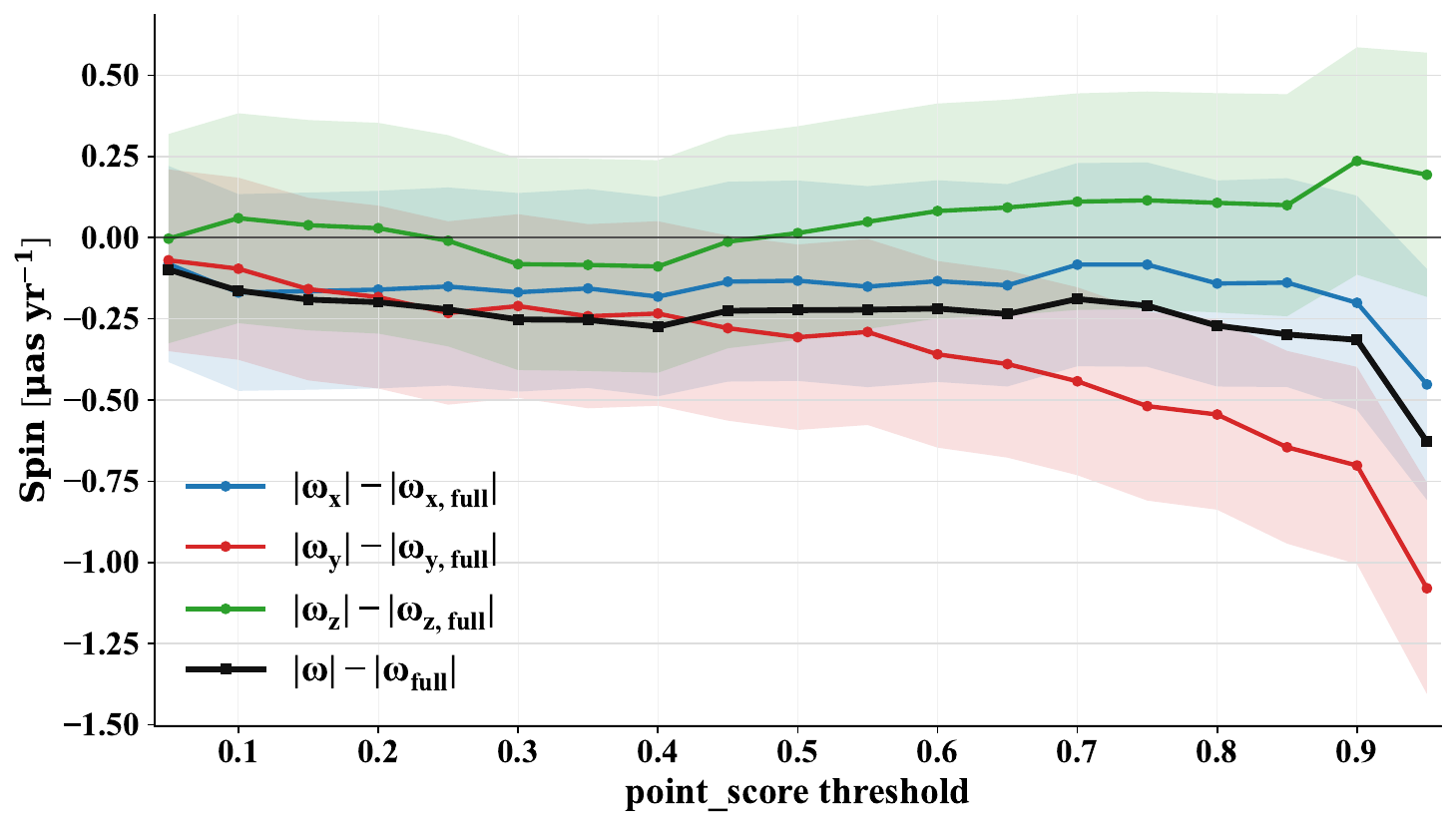}
    \caption{Change in Gaia-CRF3 spin amplitude as a function of the \texttt{point\_score} threshold. The blue, red, and green curves show the changes in the absolute spin components along the x, y, and z axes relative to the full sample, while the black curve shows the change in the total spin amplitude. Shaded regions indicate the 1$\sigma$ uncertainties of the individual spin components.
}
     \label{fig.spin}
\end{figure}

\citet{2022A&A...667A.148G} reported the spin of the full Gaia-CRF3 sample after removing about 0.4\% of extreme outliers, with values of 
\begin{equation}
\boldsymbol{\omega} =
\begin{bmatrix}
-3.44 \pm 0.30 \\
+1.57 \pm 0.28 \\
-1.24 \pm 0.32
\end{bmatrix}
\,\mu\mathrm{as}\,\mathrm{yr}^{-1}.
\end{equation}
Following the procedure described in Appendix~E of \citep{2022A&A...667A.148G}, we computed both the number of retained sources and the spin components along the three coordinate axes for different \texttt{point\_score} selection thresholds (we retained sources whose \texttt{point\_score} is greater than this threshold), after applying the same outlier rejection to each subsample. The results are presented in Table \ref{tab:total_rotation_threshold} and Fig. \ref{fig.spin}. We find that, once a \texttt{point\_score} constraint is imposed, the absolute values of the spin components along the $x$ and $y$ axes both decrease, with the effect becoming particularly clear for sources with \texttt{point\_score} $> 0.9$. By contrast, the spin component along the $z$ axis increases slightly. This suggests that the spin components along the $x$ and $y$ axes contain a contribution from morphology-related contamination, whereas the mild increase along the $z$ axis is more likely caused by the uneven sky distribution introduced by the \texttt{point\_score} selection. The total spin amplitude decreases as the \texttt{point\_score} threshold becomes more restrictive. In the most stringent case, with \texttt{point\_score} $> 0.95$, the retained sources exhibit clean point-like morphology and are largely free from contamination by nearby bright objects. The resulting total spin amplitude is $3.35~\mu{\rm as}\,{\rm yr}^{-1}$, which is 15.8\% lower than that of the full sample.

\begin{table}[htbp]
\centering
\caption{Total spin as a function of the \texttt{point\_score} threshold.}
\label{tab:total_rotation_threshold}
\begin{tabular}{ccc}
\hline
Threshold & Number of Sources & Total spin $\mu$as yr$^{-1}$ \\
\hline
0.00 & 1,607,289 & 3.9785 \\
0.05 & 1,561,272 & 3.8801 \\
0.10 & 1,542,317 & 3.8152 \\
0.15 & 1,528,914 & 3.7877 \\
0.20 & 1,517,575 & 3.7797 \\
0.25 & 1,507,192 & 3.7575 \\
0.30 & 1,497,361 & 3.7265 \\
0.35 & 1,487,308 & 3.7250 \\
0.40 & 1,476,906 & 3.7043 \\
0.45 & 1,465,421 & 3.7533 \\
0.50 & 1,452,978 & 3.7553 \\
0.55 & 1,439,139 & 3.7564 \\
0.60 & 1,423,857 & 3.7602 \\
0.65 & 1,407,000 & 3.7431 \\
0.70 & 1,385,655 & 3.7895 \\
0.75 & 1,357,984 & 3.7689 \\
0.80 & 1,319,283 & 3.7073 \\
0.85 & 1,260,546 & 3.6804 \\
0.90 & 1,150,225 & 3.6637 \\
0.95 & 819,323 & 3.3499 \\
\hline
\end{tabular}
\end{table}

\begin{figure}[ht!]
\centering
\includegraphics[width=\hsize]{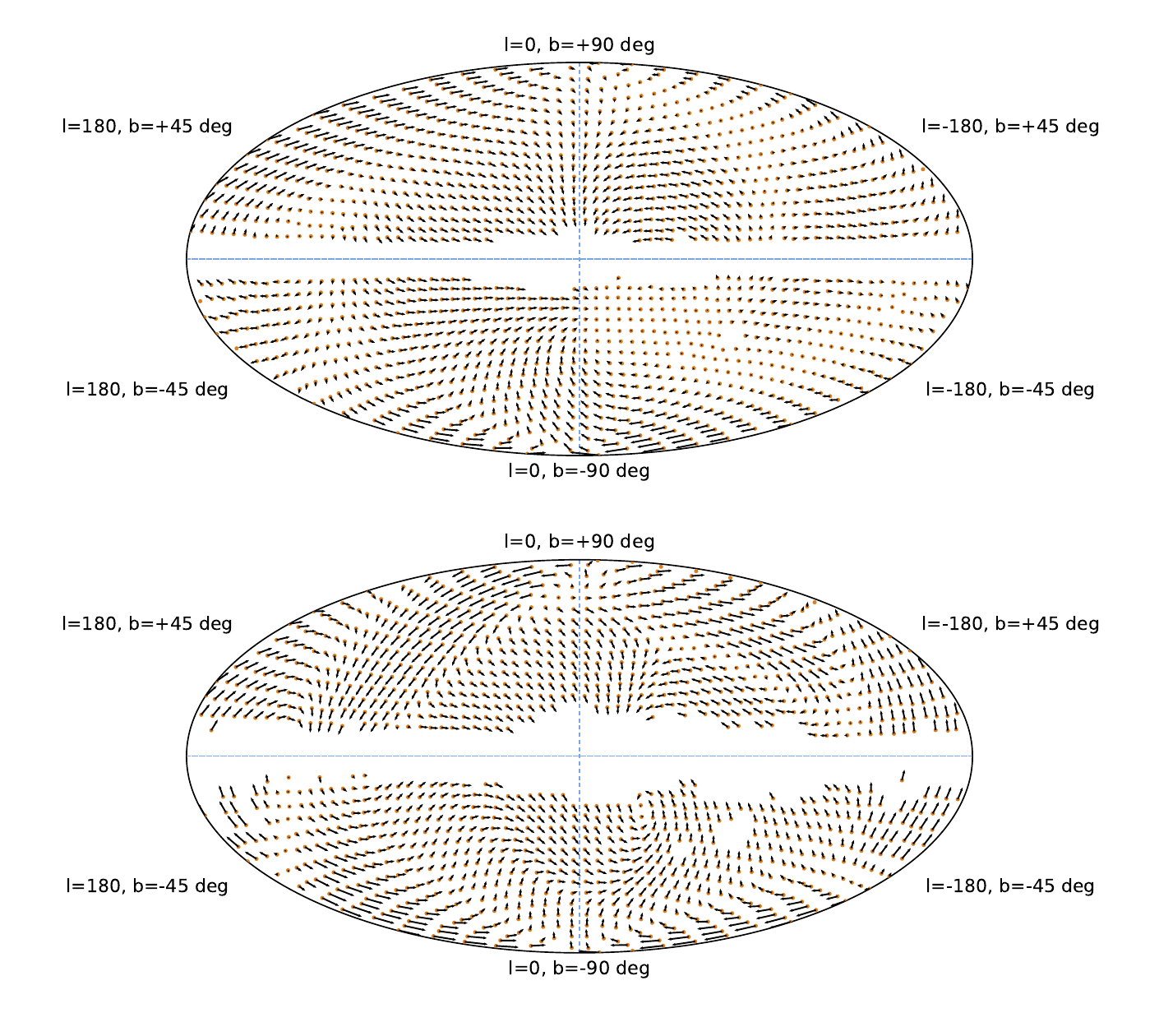}
    \caption{All-sky proper motion vector fields of Gaia-CRF3 subsamples with different morphological scores. The upper panel shows the sample with \texttt{point\_score} $> 0.95$, while the lower panel shows the sample with \texttt{point\_score} $< 0.20$. The orange points mark the centers of the pixels in Galactic coordinates, and the black arrows represent the model proper motion vectors derived from the VSH fit. Only pixels containing at least 20 sources are shown.}
     \label{fig.pm_field}
\end{figure}

Using a method similar to that adopted in \citet{2025NatAs...9.1396M}, we computed the all-sky proper motion field of two representative Gaia-CRF3 subsamples through a vector spherical harmonics (VSH, \citet{2012A&A...547A..59M}) fit, namely the sources with \texttt{point\_score} $> 0.95$ and \texttt{point\_score} $< 0.2$. These two subsamples contain 821,761 and 84,204 sources, respectively, and represent standard point-like reference sources and morphologically anomalous reference sources. The results are shown in Fig. \ref{fig.pm_field}. The high-score sample in the upper panel exhibits a smoother large-scale vector field, whereas the low-score sample in the lower panel displays a stronger and more complex large-scale pattern. We further find that the low-score sample also contains significant second- and third-order VSH terms, indicating that its proper motion field is not a purely dipolar pattern, but instead includes more complicated all-sky structure. This behaviour is consistent with the expectation that low-\texttt{point\_score} sources are more strongly affected by complex optical morphology, host-galaxy structure, source-position biases, and scanning-law residuals. This morphology-dependent difference provides an independent source-structure-based example of the hidden astrometric systematics discussed by \citet{2025NatAs...9.1396M}, who showed that VSH-fitted Gaia-CRF3 quasar proper motion fields can contain significant large-scale patterns and that such signals may be affected by residual systematic errors in Gaia astrometry. Our high- and low-\texttt{point\_score} comparison suggests that source structure is one plausible contributor to these systematics, and provides a morphology-selected reference for assessing source-structure-induced biases in future all-sky proper motion analyses. Source morphology therefore needs to be treated with particular care in future fits of all-sky proper motion fields at higher precision.

%%%%%%%%%%%%%%%%%%%%%%%%%%%%%%%%%%%%%%%%%%%%%%%%%%%%%%%%%%%%%%
\section{Discussion}
\label{sec_discussion}

\subsection{Reliability and Applicability Limits of the External Morphological Score}

The \texttt{point\_score} constructed in this work provides an optical morphological indicator for Gaia-CRF3 sources that is independent of the Gaia astrometric solution. By combining imaging information from DESI, PS1, SDSS, and SkyMapper through a multi-survey fusion scheme, this indicator reduces the impact of anomalous measurements from any individual survey. Its internal classification consistency, correlation with Gaia quality parameters, and behaviour across different magnitude ranges all show that \texttt{point\_score} is an effective tracer of the degree to which a source departs from the morphology of an ideal point source, and thus offers a new external diagnostic for assessing the quality of Gaia-CRF3 reference sources.

However, the score remains limited by the external imaging data themselves. The four surveys differ in spatial resolution, seeing, limiting magnitude, band coverage, and the definition of their morphological parameters, and these differences affect the stability of the score across sky regions and magnitude ranges. SkyMapper provides an important extension in the southern sky, but its morphological completeness and regression accuracy are lower than those of DESI and SDSS. As a result, sources for which only SkyMapper-based information is available generally have more uncertain scores. In the final catalogue, 37,346 sources rely exclusively on SkyMapper-based morphological information. Sources showing strong disagreement among the survey-based scores also require particular caution, since such discrepancies may reflect either genuine structural complexity or differences in local image quality, cross-match accuracy, and survey-specific processing.

Magnitude dependence is another important limitation. Our results show that \texttt{point\_score} remains stable for sources with $G<20.85$~mag, whereas its median value declines rapidly at fainter magnitudes. This behaviour does not necessarily imply that faint sources are intrinsically more extended, but more likely reflects the decreasing ability of the external imaging data to separate point-like and extended morphologies as the signal-to-noise ratio degrades. Therefore, when \texttt{point\_score} is applied to the faint end of Gaia-CRF3, low scores should not be interpreted directly as evidence for genuinely anomalous optical structure.

\subsection{Physical Interpretation of the Low-\texttt{point\_score} Sources}
\begin{figure}[ht!]
\centering
\includegraphics[width=\hsize]{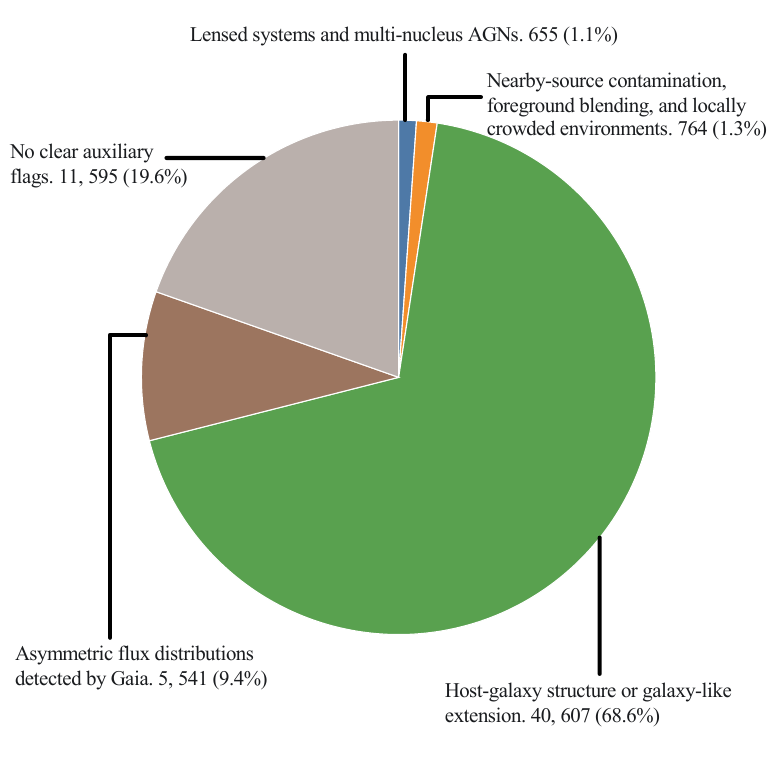}
    \caption{Pie chart showing the primary origin of the low point-source scores for sources with \texttt{point\_score} $\leq 0.1$. Each source is assigned to a single category according to the adopted hierarchical priority scheme.
}
     \label{fig.pie}
\end{figure}

Low-\texttt{point\_score} sources do not correspond to a single physical class, but rather trace the degree to which Gaia-CRF3 sources depart from the morphology of an isolated point source in external optical images. To identify the main origin of this deviation, we carried out a systematic investigation of the 59\,162 sources with \texttt{point\_score} $\leq 0.1$ by combining external catalogue cross-matches and literature-based classification. A hierarchical attribution scheme was adopted, such that each source was assigned to only one primary category. If multiple conditions were satisfied, the source was assigned to the highest-priority category. The resulting fractional distribution of the different categories is shown in Fig. \ref{fig.pie}.

\begin{enumerate}
\item \textit{Lensed systems and multi-nucleus AGNs.}
This class was identified through cross-matches with external literature catalogues, including known gravitational lenses, multiply imaged quasars, lens candidates, quasar-pair candidates, and dual- or multiple-nucleus AGN candidates \citep{2012ApJ...753...42C, 2018A&A...618A..56D, 2019MNRAS.486.4987R, 2023ApJS..264....4M, 2024A&A...685A.130G, 2025ApJS..281...25P, 2024A&A...692A.154W}. Such sources typically contain multiple nearby optical nuclei or strongly non-single-peaked light distributions, and therefore naturally deviate from the morphology of a standard point source.

\item \textit{Nearby-source contamination, foreground blending, and locally crowded environments.}
A source was assigned to this class if it has a Gaia neighbor within 3~arcsec satisfying $\Delta G \leq 4$, or if two or more Gaia neighbors are present within the same radius. Such configurations are commonly associated with local blending, foreground superposition, or crowding, all of which can significantly distort the flux distribution in the external imaging and thus shift the morphological parameters away from point-source behaviour. 

\item \textit{Host-galaxy structure or galaxy-like extension.}
This category was determined from the Gaia host-galaxy detection results, host-fitting flags, and galaxy-classification information. Sources were assigned to this class if \texttt{host\_galaxy\_detected = true}, if \texttt{host\_galaxy\_flag} indicated a host-galaxy detection or an anomalous host fit, or if Gaia galaxy-candidate flags and galaxy-classification probabilities suggested an overall galaxy-like morphology. In these cases, the low score is primarily associated with resolved extension intrinsic to the source itself rather than with local blending.

\item \textit{Asymmetric flux distributions detected by Gaia.}
Gaia-based image diagnostics provide an independent line of evidence for morphology-related anomalies. Sources were assigned to this class if they show BP/RP window blending or contamination, excess BP/RP flux relative to the $G$ band, multi-peaked image-parameter solutions, or auxiliary flags such as odd-window or duplicate-source indicators. In such cases, Gaia itself already detects an asymmetric flux distribution inconsistent with an isolated point source.

\item \textit{No clear auxiliary flags.}
All remaining low-score sources not covered by the criteria above were assigned to this class. These objects show some degree of extension in the external imaging, but lack direct support from Gaia auxiliary parameters or other external survey flags. They may include weakly extended systems not captured by the current diagnostics, but may also be affected by image quality, local background structure, or cross-match uncertainties.
\end{enumerate}

As a comparison sample, we applied the same attribution scheme to sources with \texttt{point\_score} $>0.9$. The fractions of the main physically interpretable contaminating classes are strongly reduced in this high-score regime, with only 0.13\% classified as lensed systems or multi-nucleus AGNs, 0.24\% as nearby-source contamination or crowded environments, and 0.14\% as host-galaxy structure or galaxy-like extension.

We further examined the relation between \texttt{point\_score} and redshift using the photometric redshifts \texttt{z\_ph} from CatGlobe \citep{2025ApJS..279...54F}. The score shows a positive trend with redshift, in the sense that higher-redshift sources tend to have larger \texttt{point\_score} values. Conversely, the low-score population is strongly concentrated at low redshift. Among sources with \texttt{point\_score} $\leq 0.1$ and available CatGlobe \texttt{z\_ph}, 81.0\% have $z_{\rm ph}<0.5$, and 97.4\% have $z_{\rm ph}<1$. This redshift dependence is consistent with the interpretation that a substantial fraction of low-\texttt{point\_score} sources is affected by resolved host-galaxy structure, nearby blending, or other extended optical morphology that is more readily detected at low redshift. Together with the redshift dependence described above, the pie chart in Fig. \ref{fig.pie} shows that the origin of low-\texttt{point\_score} sources is intrinsically diverse. In addition to a minority of systems that can be directly explained by known lenses, dual nuclei, or quasar pairs, a substantial fraction of low-score sources is associated with nearby-source contamination, crowded environments, or resolved host-galaxy structure. Another fraction is characterised mainly by asymmetric flux distributions already detected by Gaia or by extension visible only in the external imaging. This indicates that \texttt{point\_score} identifies a broader population of non-ideal point-like reference sources.

\subsection{Prospects for future reference-frame construction and quasar studies}

The most direct application of \texttt{point\_score} is to improve source selection for optical celestial reference frames. The construction of Gaia-CRF3 mainly emphasized sample purity, sky uniformity, and the statistical properties of the astrometric parameters, whereas optical morphology has not yet been incorporated systematically as an independent variable in source quality control. Our results show that low-\texttt{point\_score} sources introduce larger parallax zero-point residuals and stronger proper motion systematics, while high-\texttt{point\_score} sources define a more stable reference source subsample. Morphological information should therefore be treated as an important input for source selection, weight assignment, and systematic-error modeling in future realizations such as Gaia-CRF4 and other high-precision optical reference frames.

The \texttt{point\_score} is also of practical value for parallax zero-point correction and all-sky proper motion field fitting. Current Gaia parallax zero-point models mainly depend on variables such as magnitude, colour, and ecliptic latitude, whereas our results show that morphology-related residuals remain significant in low-score sources. For studies that use large AGN samples to fit frame spin and higher-order vector spherical harmonics, morphological selection can reduce the contamination of the large-scale proper motion field by non-point-like sources and thereby improve the reliability of the inferred physical signals.

Low-\texttt{point\_score} sources are themselves of considerable scientific interest. They are not only contaminants to be excluded from reference-frame construction, but may also include quasars with prominent host galaxies, dual AGNs, lensed AGNs, and systems with anomalous photocenter variability. Combined with significant parallaxes, significant proper motions, large \texttt{astrometric\_excess\_noise}, multi-component image properties, and multi-band photometric anomalies, \texttt{point\_score} may provide an efficient way to identify such candidates. These samples can in turn be used to study AGN host-galaxy structure, galaxy mergers, strong gravitational lenses, and the response of Gaia photocenter measurements to complex extragalactic sources.

%%%%%%%%%%%%%%%%%%%%%%%%%%%%%%%%%%%%%%%%%%%%%%%%%%%%%%%%%%%%%%

\section{Conclusions}
\label{sec_conclusion}
As the most accurate celestial reference frame currently available in the optical band, Gaia-CRF3 provides not only the fundamental spatial reference for modern astrometry, but also an essential benchmark for assessing systematic errors in astrometric catalogues. The origin of the systematics in Gaia-CRF3 is intrinsically complex, as it couples the physical properties of the sources with instrumental effects and with the data processing and calibration pipelines. Identifying these systematics and mitigating them is therefore crucial for improving the reliability of the reference frame.

In this work, we assessed the morphology of 1\,607\,490 Gaia-CRF3 reference sources, corresponding to a coverage of 99.59\%, using morphological parameters derived from the imaging data of the DESI Legacy Imaging Surveys, SDSS, and SkyMapper, together with the morphological scores from PS1-PSC. We first derived independent morphology scores from each external survey and performed internal consistency tests. The scores based on DESI and SDSS show high consistency with their native morphological classifications. Although the consistency for SkyMapper is slightly lower, its AUC still reaches 0.94, indicating that these scores effectively trace the source morphology within each survey.

To improve the usability of the results, we developed a fusion strategy that combines the information from different surveys and assigns a single parameter, \texttt{point\_score}, to each Gaia-CRF3 source. This parameter quantifies the degree to which a source deviates from a standard point-like morphology. Several Gaia DR3 parameters, such as \texttt{astrometric\_excess\_noise} and \texttt{phot\_bp\_rp\_excess\_factor}, are expected to be affected by source morphology. We therefore examined the correlations between \texttt{point\_score} and these two Gaia parameters. Both comparisons reveal clear monotonic trends, demonstrating that Gaia parameters can indirectly encode morphological information. This is consistent with the known sensitivity of Gaia astrometric and photometric diagnostics to source structure, blending, and extended morphology.

We find that the median \texttt{point\_score} remains above 0.9 and shows no significant fluctuation for sources brighter than \(G=20.85\) mag. For fainter sources, however, the scores become less reliable, primarily because of the limiting magnitudes of the external imaging surveys. A study of the normalised astrometric parameters in different \texttt{point\_score} intervals shows that, within each magnitude bin, sources with higher scores have distributions that are closer to a standard Gaussian. This indicates that source morphology has a significant impact on astrometric reliability: samples with more point-like morphologies exhibit smaller systematic errors.We further find that low-\texttt{point\_score} sources have a raw weighted mean parallax of \(-43.7~\mu\mathrm{as}\) in the range \(18 \leq G < 20.5\). After applying the parallax zero-point correction model, these sources still retain a corrected residual of \(-26.9~\mu\mathrm{as}\), corresponding to a residual difference of \(28.2~\mu\mathrm{as}\) relative to the high-score sample. This effect is mainly associated with six-parameter solutions. We find that the all-sky proper motion fields differ significantly among subsamples defined by different \texttt{point\_score} ranges. Moreover, applying a stringent \texttt{point\_score} selection reduces the spin amplitude of Gaia-CRF3 subsamples. For the subsample with \texttt{point\_score} $> 0.95$, the total spin amplitude is reduced by 15.8\% relative to that of the full sample.

The construction of existing celestial reference frames has carefully considered the reliability of the sources as AGN and the distribution properties of the sample's normalised astrometric parameters. However, the impact of source morphology on the astrometric parameters has not yet been explicitly taken into account. Our results indicate that morphology-related information should be incorporated as an additional quality-control indicator in future Gaia releases, for the screening and validation of astrometric solutions. Based on our analysis, we conclude that AGN whose morphology strongly deviates from that of a point source should not be used to construct celestial reference frames at sub-mas accuracy. Such sources, especially those with \texttt{point\_score} $\leq 0.2$, should be identified and removed from current realisations where possible, and treated with caution in future reference-frame construction.

Beyond its application to the construction of a more reliable celestial reference frame, \texttt{point\_score} also provides a useful morphological diagnostic for quasar samples. It may be used in future studies to search for AGN with prominent host-galaxy contributions, as well as sources that deviate from standard Gaussian astrometric behaviour, such as dual AGN and lensed quasars.

%%%%%%%%%%%%%%%%%%%%%%%%%%%%%%%%%%%%%%%%%%%%%%%%%%%%%%%%%%%%%%
\section*{Data availability}
\label{sec_data}
The \texttt{point\_score} catalogue will be made publicly available through the CDS upon acceptance of this paper.

%%%%%%%%%%%%%%%%%%%%%%%%%%%%%%%%%%%%%%%%%%%%%%%%%%%%%%%%%%%%%%

%%%%%%%%%%%%%%%%%%%%%%%%%%%%%%%%%%%%%%%%%%%%%%%%%%%%%%%%%%%%%%

\begin{acknowledgements}
    This work was supported by the National Key R\&D Program of China (Grant No.2025YFA1614104), the National Natural Science Foundation of China (NSFC) through grants 12173069, the Strategic Priority Research Program of the Chinese Academy of Sciences, Grant No.XDA0350205, the Youth Innovation Promotion Association CAS with Certificate Number 2022259, the Talent Plan of Shanghai Branch, Chinese Academy of Sciences with No.CASSHB-QNPD-2023-016, the International Partnership Program of the Chinese Academy of Sciences with Grant No.018GJHZ2025032FN. This work has made use of data from the European Space Agency (ESA) mission Gaia, DESI Legacy Imaging Surveys, SkyMapper and Sloan Digital Sky Survey (SDSS). We are also very grateful to the developers of the TOPCAT \citep{2005ASPC..347...29T} software, the SIMBAD astronomical database and HEALPix. 
\end{acknowledgements}

% Please note that we have included the references below in
% order to compile the document, but we ask you to:
\bibliographystyle{aa} % style aa.bst
\bibliography{cite_paper} % your references Yourfile.bib

%%%%%%%%%%%%%%%%%%%%%%%%%%%%%%%%%%%%%%%%%%%%%%%%%%%%%%%%%%%%%%%
% Appendices must be placed after   \end{thebibliography}
% They will be placed automatically on a new page.
%%%%%%%%%%%%%%%%%%%%%%%%%%%%%%%%%%%%%%%%%%%%%%%%%%%%%%%%%%%%%%%
\begin{appendix}

\section{normalised astrometric distributions in Morphological Score and magnitude bins}
\label{sec:normalised}

\begin{figure*}[!t]
\centering
\includegraphics[width=\hsize]{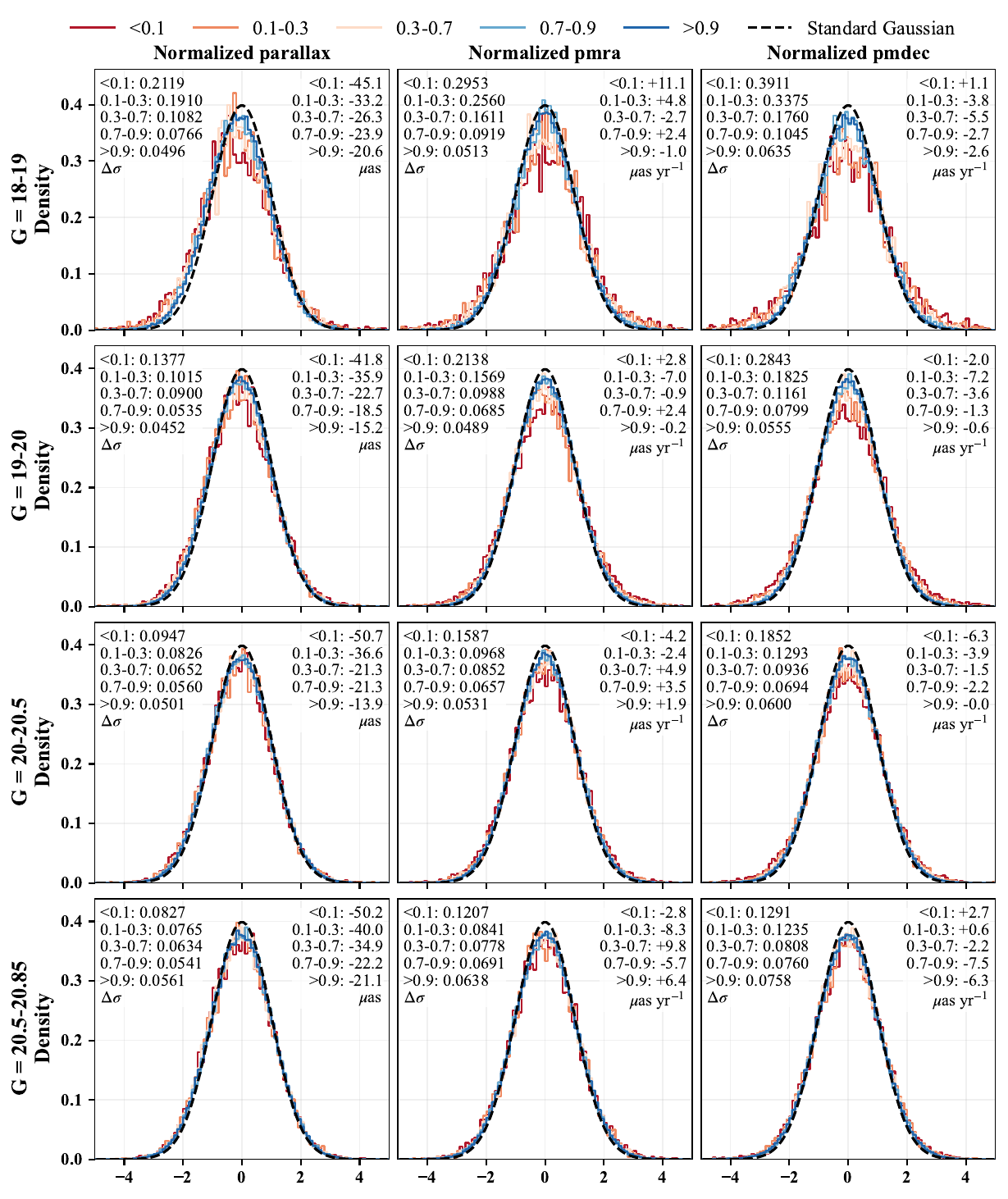}
\caption{Normalised astrometric distributions in four $G$-magnitude bins and five \texttt{point\_score} bins. Rows correspond to magnitude bins, columns show $\varpi/\sigma_{\varpi}$, $\mu_{\alpha *}/\sigma_{\mu_{\alpha *}}$, and $\mu_{\delta}/\sigma_{\mu_{\delta}}$, and the dashed black curve shows the standard normal distribution. The numbers in the upper-left corner of each panel give $\Delta\sigma=\sigma_{\rm fit}-1$ for the best-fit Gaussian standard deviation of each \texttt{point\_score} subsample. The numbers in the upper-right corner of each panel give the weighted means of the corresponding raw parallax or proper-motion component for the five \texttt{point\_score} subsamples, expressed in $\mu$as for parallax and $\mu$as\,yr$^{-1}$ for proper motion.}
     \label{fig.astrometry_guss}
\end{figure*}

Fig. \ref{fig.astrometry_guss} shows the distributions of the normalised Gaia astrometric quantities
$\varpi/\sigma_{\varpi}$,
$\mu_{\alpha *}/\sigma_{\mu_{\alpha *}}$,
and
$\mu_{\delta}/\sigma_{\mu_{\delta}}$
for the Gaia-CRF3 sources with valid \texttt{point\_score} and
$18 \leq G < 20.85$.
The four rows correspond to the magnitude bins
$18$--$19$,
$19$--$20$,
$20$--$20.5$,
and
$20.5$--$20.85$,
while the three columns show the normalised parallax, normalised pmra, and normalised pmdec distributions, respectively. Within each panel, the coloured histograms represent the five \texttt{point\_score} bins, and the dashed black curve gives the standard normal distribution.
This figure provides a direct visual complement to the KS-distance and tail-excess statistics discussed in Section \ref{cross_survey}.

At fixed magnitude, the high-score subsamples are systematically closer to the standard normal profile, with narrower wings and weaker non-Gaussian tails, whereas the low-score subsamples show broader distributions, especially in the brighter magnitude bins and in the proper motion components. Toward fainter magnitudes, the contrast between score bins becomes smaller, which is consistent with the increasing contribution of formal measurement noise.

\section{Workflow for Morphological Parameter Selection and Score Regression}
\label{sec:workflow}
\begin{figure}[ht!]
\centering
\includegraphics[width=\hsize]{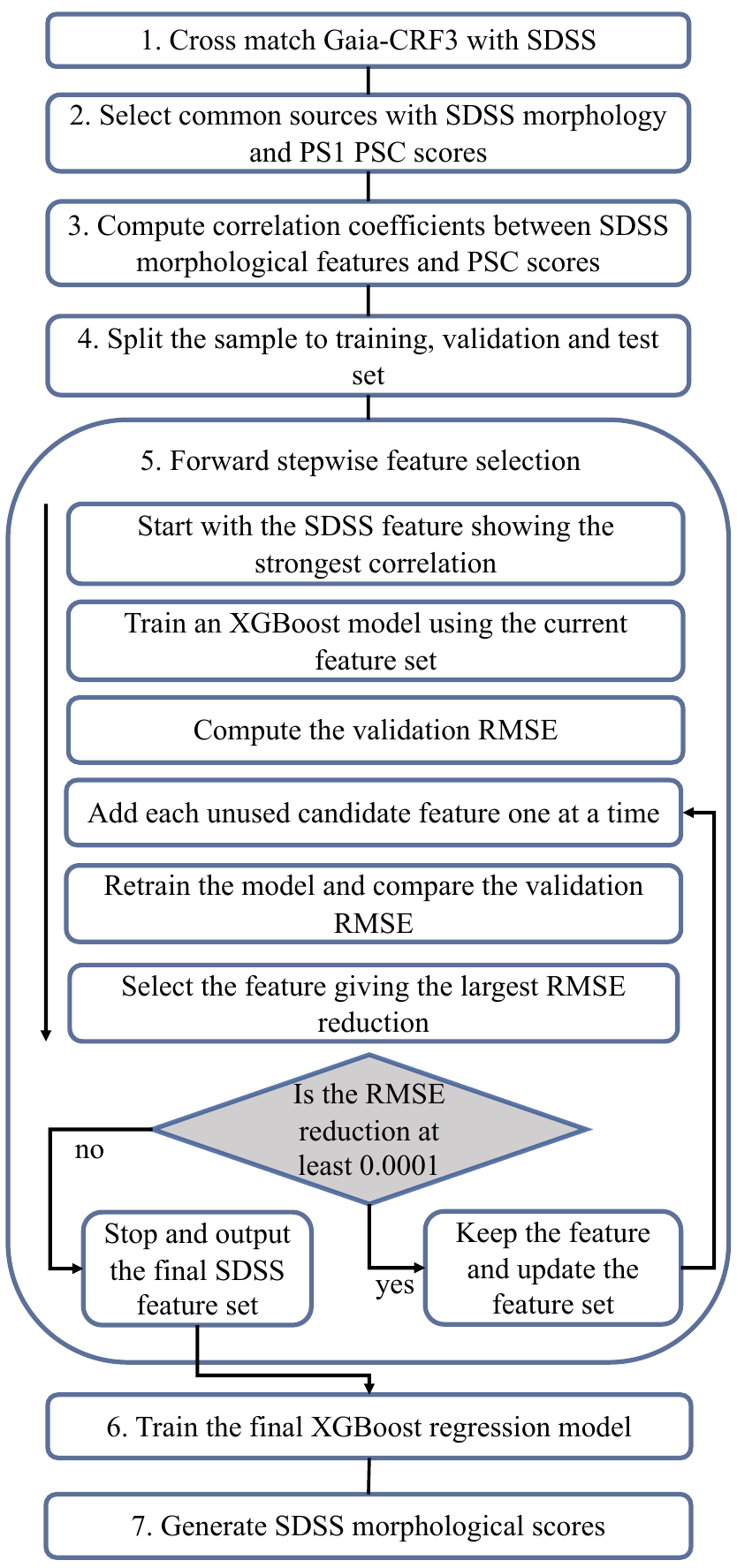}
    \caption{Workflow of the SDSS-based morphological score construction.
}
     \label{fig:workflow}
\end{figure}

This appendix describes the procedure used to select the morphological input parameters and to construct survey-dependent morphological scores on the PS1-PSC score scale. The same general strategy was adopted for SDSS, DESI, and SkyMapper, while Fig.~\ref{fig:workflow} illustrates the procedure using SDSS as an example.

For each external imaging survey, Gaia-CRF3 sources with both survey-specific morphological measurements and PS1-PSC scores were used as the supervised sample, with the PS1-PSC score adopted as the reference scale. The candidate morphological and quality-related parameters were first ranked according to their Pearson and Spearman correlations with the PS1-PSC score. The sample was then split, with stratification in PS1-PSC score, into training and validation, and test sets with fractions of 0.70, 0.15, and 0.15, respectively. The final input set was determined through a forward stepwise procedure: starting from the most strongly correlated parameter, each unused candidate was added in turn, an XGBoost regression model was retrained, and the validation-set RMSE was recomputed. At each iteration, the parameter yielding the largest RMSE reduction was retained, and the procedure was stopped when the improvement became smaller than 0.0001. The selected parameters were then used to train the final XGBoost model, whose performance was evaluated on the independent test set using the RMSE, MAE, and Pearson correlation coefficient before being applied to all matched sources to generate the survey-specific morphological scores.

\section{Multi-survey Fusion Strategy for the Composite Morphological Score}
\label{sec:fusion}

The composite morphological score is designed to combine the survey-based point-source scores from DESI, PS1, SDSS, and SkyMapper onto a common scale. A hard-priority rule, such as ${\rm DESI} > {\rm PS1} > {\rm SDSS} > {\rm SkyMapper}$, is simple, but visual inspection showed that it can occasionally amplify anomalous measurements from a single survey, especially when DESI assigns a high score to a source for which the other available surveys consistently give much lower values. We therefore adopted a weighted multi-survey fusion scheme in which DESI remains the most influential input, but no longer has absolute priority.

For each source, we denote the available survey-based scores by $s_{\rm D}$, $s_{\rm P}$, $s_{\rm S}$, $s_{\rm K}$, for DESI, PS1, SDSS, and SkyMapper, respectively, with $0 \leq s_i \leq 1$. The adopted base weights are
\begin{equation}
w_{\rm D}=0.45,\quad
w_{\rm P}=0.30,\quad
w_{\rm S}=0.15,\quad
w_{\rm K}=0.10.
\end{equation}
For a given source, only valid scores are used. If the set of available surveys is
\begin{equation}
A=\{i \mid s_i\ {\rm is\ available}\},
\end{equation}
the renormalised weights are
\begin{equation}
\tilde{w}_i=\frac{w_i}{\sum_{j\in A}w_j}, \qquad i\in A.
\end{equation}

When at least two surveys provide valid scores, we first compute the weighted mean
\begin{equation}
S_{\rm mean}=\sum_{i\in A}\tilde{w}_i s_i,
\end{equation}
and the inter-survey disagreement
\begin{equation}
D_{\rm survey}=
\sqrt{
\sum_{i\in A}\tilde{w}_i
\left(s_i-S_{\rm mean}\right)^2
}.
\end{equation}
The default fused score is then defined as
\begin{equation}
S_{\rm final}=S_{\rm mean}-\lambda D_{\rm survey},
\end{equation}
with $\lambda=0.3$, this value was chosen empirically from the distribution of $D_{\rm survey}$ for the 1,452,804 sources with non-null inter-survey scatter. The 90th, 95th, and 99th percentiles of $D_{\rm survey}$ are 0.123, 0.193, and 0.300, respectively, which correspond to score corrections of approximately 0.04, 0.06, and 0.09 for $\lambda=0.3$. The adopted penalty therefore leaves the majority of sources nearly unchanged, while still providing a meaningful down-weighting for the most discrepant cases. The resulting score is finally truncated to the interval $[0,1]$.

A special treatment is applied when DESI yields a high score but the other surveys consistently support a much lower value. If DESI is available and at least two additional surveys also provide valid scores, we define
\begin{equation}
S_{\rm other}=
{\rm median}
\left(
s_i,\ i\in A,\ i\neq {\rm D}
\right).
\end{equation}
If $s_{\rm D}>0.8$, $S_{\rm other}<0.6$, and $s_{\rm D}-S_{\rm other}>0.3$, the source is flagged as \texttt{DESI\_HIGH\_CONFLICT} and the final score is conservatively set to $S_{\rm final}=S_{\rm other}$.

If only one survey provides a valid score, that score is directly adopted and the source is flagged as \texttt{SINGLE\_SURVEY}. If no score is available, the final score is set to \texttt{NaN}. The full definition is therefore
\begin{equation}
S_{\rm final}=
\begin{cases}
\texttt{NaN}, & N_{\rm survey}=0, \\[4pt]
s_i, & N_{\rm survey}=1, \\[4pt]
S_{\rm other}, & {\rm DESI\_HIGH\_CONFLICT}, \\[4pt]
S_{\rm mean}-\lambda D_{\rm survey}, & {\rm otherwise}.
\end{cases}
\end{equation}

This fusion scheme is more robust than a hard-priority assignment and reduces the impact of survey-specific outliers on the final morphological score.

\end{appendix}
\end{document}